\documentclass[prx,superscriptaddress]{revtex4-1}

\usepackage{amsmath}
\usepackage{graphicx}
\usepackage{smallgrf}
\usepackage{amssymb}
\usepackage{mathrsfs}
\usepackage{bm}

\newtheorem{Definition}{Definition.}

\begin{document}

\title{Conservation law in noncommutative geometry\\
-- Application to spin-orbit coupled systems}
\author{Naoyuki Sugimoto}
\affiliation{Cross-correlated Materials Research Group (CMRG) 
and Correlated Electron Research Group (CERG), 
RIKEN, Saitama 351-0198, Japan}
\author{Naoto Nagaosa}
\affiliation{Cross-correlated Materials Research Group (CMRG) 
and Correlated Electron Research Group (CERG), 
RIKEN, Saitama 351-0198, Japan}
\affiliation{Department of Applied Physics University of Tokyo, 
Tokyo 113-8656, Japan}

\begin{abstract}
The quantization scheme by noncommutative geometry developed in 
string theory is applied to establish the conservation law 
of twisted spin and spin current densities in the
spin-orbit coupled systems. Starting from the pedagogical 
introduction to Hopf algebra and deformation quantization, 
the detailed derivation of the conservation law is given. 
\end{abstract}
\maketitle
\tableofcontents

\clearpage

\section{Introduction}

Electrons are described by the Dirac equation where the U(1) Maxwell 
electromagnetic field (emf) $A_\mu$ is coupled to the charge current $j_\mu$ 
as described by the Lagrangian
(in the natural unit where $\hbar = c =1$; $\mu=0,1,2,3$)~\cite{Peshkin}
\begin{equation}
L = {\bar \psi} [{\rm i}\gamma^\mu {\hat D}_\mu - m ] \psi.
\end{equation}
where ${\hat D}_\mu = \partial_\mu - {\rm i}e A_\mu$ 
is the covariant derivative, $m$ is the electron mass.
Note that the spin is encoded by 4 component nature of the spinors $\psi$
and ${\bar \psi} = \psi^\dagger \gamma^0$
and the $4\times 4$ gamma matrices $\gamma^\mu$, 
but the charge and charge current alone determine 
the electromagnetic properties of the electrons, which
are given by 
\begin{equation}
 j^\mu = - { {\partial  L} \over {\partial A_\mu} } = 
 -e {\bar \psi} \gamma^\mu \psi.
\end{equation}

In condensed matter physics, 
on the other hand, the low energy phenomena compared 
with the mass gap $2mc^2 \sim 10^6 eV$ are considered, 
and only the positive energy states described by the two-component 
spinor are relevant.  Then, the relativistic spin-orbit interaction 
originates when the negative energy states (positron stats) are projected 
out to derive the effective Hamiltonian or Lagrangian. 
The projection to a subspace of the Hilbert space leads 
to the nontrivial geometrical structure which is 
often described by the gauge theory. 
This is also the case for the Dirac equation, 
and the resultant gauge field is SU(2) 
non-Abelian gauge field corresponding to the Zeeman effect (time-component) 
and the spin-orbit interaction (spatial components) as described below.

The effective Lagrangian for the positive energy states
can be derived by the expansion with respect to 
$1/(mc^2)$~\cite{Froelich,He,Zaanen}
\begin{equation}
 L = {\rm i} \psi^\dagger D_0 \psi   
+ \psi^\dagger  { { {\bm D}^2} \over { 2m}}  \psi +
{ 1 \over {2m}} \psi^\dagger \biggl[
eq \sigma^a {\bm A} \cdot {\bm A}^a + { {q^2} \over 4} {\bm A}^a \cdot 
{\bm A}^a \biggr] \psi,
\label{eq:Lag}
\end{equation}
where $\psi$ is now the two-component spinor and 
$D_0 = \partial_0 + {\rm i}e A_0 + {\rm i}q A_0^a { {\sigma^a} \over 2}$,
and $D_i = \partial_i - {\rm i}e A_i - {\rm i}q A_i^a { {\sigma^a} \over 2}$ $(i=1,2,3)$
are the gauge covariant derivatives with $q$ being the quantity
proportional to the Bohr magneton~\cite{Froelich,Zaanen}. 
$A_\mu$ is the Maxwell emf, and the SU(2) gauge
potential are defined as
\begin{eqnarray}
A_0^a &=& B_a
\nonumber \\
A_i^a &=& \epsilon_{ia \ell} E_{\ell},
\end{eqnarray}
and $\sigma^{x,y,z}$ represent the Pauli matrices. 
The SU(2) gauge field is coupled to 
the 4-component spin current 
\begin{eqnarray}
j^a_0 &=& \psi \sigma^a \psi,
\nonumber \\ 
j^a_i &=&  { 1 \over { 2 m {\rm i}}}
[ \psi^\dagger \sigma^a D_i  \psi - D_i \psi^\dagger \sigma^a \psi ].
\end{eqnarray}
Namely, the Zeeman coupling and the spin-orbit interaction 
can be regarded as the gauge coupling between the 
4-spin current and the SU(2) gauge potential.
(The spin current is the tensor quantity with one suffix for the
direction of the spin polarization while 
the other for the direction of the flow.)
Note that the system has no SU(2) gauge symmetry 
since the ``vector potential'' $A^a_\mu$ is given by the physical 
field strength ${\bm B}$ and ${\bm E}$, i.e., the relation 
$\partial^\mu A^a_\mu=0$ automatically holds.    
This fact is connected to the absence of the 
conservation law for the spin density and spin current density 
in the presence of the relativistic spin-orbit interaction. 
In the spherically symmetric systems, the total angular momentum,
i.e., the sum of the orbital and spin angular momenta, is conserved, but
the rotational symmetry is usually broken by the periodic or disorder
potential $A_0$ in condensed matter systems. Therefore, it is usually assumed
that the conservation law of spin is lost by the spin-orbit interaction.

However, it is noted that the spin and spin current densities are
``covariantly'' conserved as described by the 
``continuity equation''~\cite{Froelich,He,Zaanen}
\begin{equation}
D_0 J^a_0 + {\bm D} \cdot {\bm J}^a = 0.
\end{equation}
replacing the usual derivative $\partial_\mu$ 
by the covariant derivative $D_\mu$.
This suggest that the conservation law holds in the
co-moving frame, but the crucial issue is how to 
translate this law to the laboratory frame, which is
the issue addressed in this paper.
Note again that the SU(2) gauge symmetry is absent 
in the present problem, and hence the Lagrangian 
like ${\rm tr}(F_{\mu \nu} F^{\mu \nu})$, which usually
leads to the generalized Maxwell equation and also
to the conservation law of 4 spin current including both
the matter field and gauge field~\cite{Peshkin}, is missing.
Instead, we will regard $A^a_\mu$ as the frozen
background gauge field, and focus on the 
quantum dynamics of noninteracting electrons only.

In this paper, we derive the hidden conservation law
by defining the ``twisted'' spin and spin current densities 
which satisfy the continuity equation with 
the usual derivative $\partial_\mu$.
The description is intended to be pedagogical and
self-contained. 
For this purpose, the theoretical techniques developed in 
high energy physics is useful. 
The essential idea is to take into account the effect of 
the background gauge field
in terms of the noncommutative geometry generalizing the
concept of ``product''. 
This is achieved by extending the usual Lie algebra to Hopf algebra. 

Usually, a conservation law is derived from symmetry of an action,
i.e., Noether's theorem. 
The symmetry in the noncommutative geometry is called as a ``twisted''
symmetry, and this symmetry and the corresponding 
generalized Noether's theorem have been studied in the high energy physics. 
Seiberg and Witten proposed that an equivalence of a certain string
theory and a certain field theory in noncommutative
geometry~\cite{SeibergWitten}. 
Since then, the noncommutative geometry have been attracted many researchers. 
On the other hand, it is known that the Poincare symmetry is broken in
a field theory on a noncommutative geometry. 
It is a serious problem because the energy and momentum cannot be defined. 
M. Chaichian, {\it et al}. proposed the twisted symmetry 
in the Minkowski spacetime, 
and alleged that the twisted Poincare symmetry is substituted for 
the Poincare symmetry~\cite{Chaichian200498,PhysRevLett.94.151602}. 
Moreover, G. Amelino-Camelia, {\it et al}. discussed Noether's theorem in 
the noncommutative geometry~\cite{PTPS.171.65,AmelinoCamelia2009298}. 

As we will discuss in detail later, 
a certain type of a noncommutative geometry space is equal to 
a spin-orbit coupled system. 
Therefore, a global SU(2) gauge symmetry in the noncommutative geometry
space gives a Noether current corresponding to the 
``twisted'' spin and spin current in the spin-orbit coupling system. 
This enables us to derive the generalized Noether's theorem for 
the twisted spin and spin current densities.  

Now some remarks about the application is in order.
Spintronics is an emerging field of electronics where the role of charge 
and charge current are replaced by the spin and spin current aiming 
at the low energy cost functions~\cite{Maekawa}. 
The relativistic spin-orbit interaction 
plays the key role there since it enables the manipulation of spins by the
electric field. However, this very spin-orbit interaction introduces the
spin relaxation which destroys the spin information in sharp contrast to
the case of charge where the information is protected by the
conservation law. Therefore, it has been believed that the
spintronics is possible in a short time-scale or the small size
devices. The discovery of the conservation law of twisted spin 
and spin current densities means that the quantum information of spin 
is preserved by this hidden conservation law, and could be recovered.
Actually, it has been recently predicted that the adiabatic change
in the spin-orbit interaction leads to the recovery of the spin moment 
called spin-orbit echo~\cite{Sugimoto2}. Therefore, the conservation law
of the twisted spin and spin current densities is directly related to the
applications in spintronics. 

\begin{figure}[t]
\begin{center}
\includegraphics[width=12cm,clip]{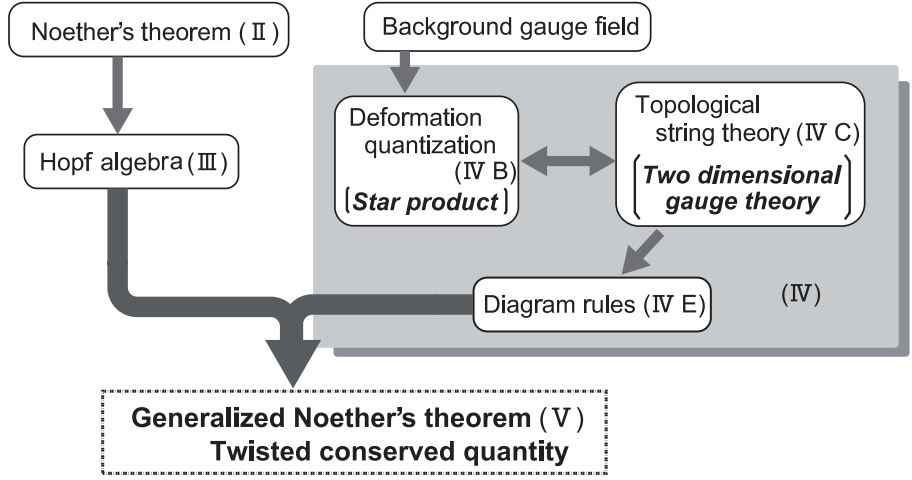}
\end{center}
\caption{
Flows of derivation of generalized Noether's theorem. 
Roman numerals and capital letters in boxes represent 
section and subsection numbers, respectively. 
A generalization of the Noether's theorem is achieved 
through Hopf algebra and deformation quantization (section V). 
Hopf algebra appear to characterize feature of an infinitesimal
transformed variation operator (sections I\!\,I and I\!\,I\!\,I). 
The SU(2) gauge structure is embedded in the star product 
(section I\!\,V).
}
\label{fig:flow}
\end{figure}

The plan of this paper follows (see Fig.~\ref{fig:flow}). 
In section \ref{sec:Noether}, we review the conventional 
Noether's theorem, and describe briefly
its generalization to motivate the use of Hopf algebra 
and deformation quantization.
In section \ref{sec:Hopf}, the Hopf algebra 
is introduced, and section \ref{sec:star} gives the explanation 
of the deformation quantization with the star product. 
The gauge interaction is compactly taken into account
in the definition of the star product. 
These two sections are sort of short review for the 
self-containedness and do not contain any 
original results except the derivation of the 
star product with gauge interaction.
Section \ref{section:TS} is the main body of this paper.
By combining the Hopf algebra and the
deformation quantization, we present the
derivation of the conserved twisted spin and spin current densities.
Section \ref{sec:Conc} is a brief summary of the paper and contains the 
possible new directions for future studies. 
The readers familiar with the
noncommutative geometry and deformation quantization can skip
sections \ref{sec:Hopf}, \ref{sec:star}, 
and directly go to section \ref{section:TS}. 

\section{Noether's theorem in field theory}\label{sec:Noether}

In this section, we discuss Noether's 
theorem~\cite{ConservationTheorems}, and its generalization
as a motivation to introduce the Hopf algebra and deformation
quantization. 
In section \ref{subsec:con}, 
we will recall Noether theorem, and rewrite it using 
the so-called ``coproduct'', which is an element of the Hopf algebra. 
In section \ref{subsec:general}, 
we will sketch a derivation of generalized Noether theorem. 

\subsection{Conventional formulation of Noether's theorem}\label{subsec:con}

We start with the action $I$ given by  
\begin{eqnarray}
I&=&\int_\Omega d^{\rm Dim}x{\cal L}(x)\nonumber\\
&=&\int d^{\rm Dim}xh_\Omega(x){\cal L}(x),
\end{eqnarray}
where $\Omega$ represents a range of the spacetime coordinate $x$ 
($\equiv(x^0,x^i)\equiv(ct,\bm{x})$) with a dimension Dim, i.e., 
$({\rm Dim}-1)$ is the dimension of the space, 
${\mathcal L}$ describes a Lagrangian density, and
\begin{equation}
 h_\Omega(x)=\left\{
\begin{array}{ccc}
1&{\rm for}&\{x|x\in\Omega\}\\
0&{\rm for}&\{x|x\notin\Omega\}.
\end{array}
\right.;
\end{equation}
$c$ represents light speed. 
We introduce a field $\phi_r$ with internal degree of freedom $r$,
and infinitesimal transformations: 
\begin{eqnarray}
 &&x^\mu\mapsto({x^\prime})^\mu:=x^\mu+\delta_\zeta x^\mu,\label{eq:TFx}\\
 &&\phi_r(x)\mapsto\phi^\prime_r(x^\prime)
:=\phi_r(x)+\delta_\zeta \phi_r(x),\label{eq:TFf}
\end{eqnarray}
where we characterize the transformations by the subscript; 
specifically, $\zeta$ represents a general infinitesimal transformation. 
Hereafter, we will employ Einstein summation convention, i.e.,
$a^\mu b_\mu\equiv a_\mu b^\mu\equiv
\sum_{\mu=0}^{{\rm Dim}-1}\eta_{\mu\nu}a^\mu b^\mu$ 
with vectors $a^\mu$ and $b^\mu$ 
$(\mu=0,1,\ldots,({\rm Dim}-1))$, 
and the Minkowski metric: 
$\eta_{\mu\nu}:={\rm diag}(-1,\underbrace{1,1,\ldots, 1}_{{\rm Dim}-1})$. 

We define the variation operator of the action as follow:
\begin{eqnarray}
\delta_\zeta I&:=&
\int_{\Omega^\prime}{d^{\rm Dim}x^\prime}{\cal L}^\prime({x^\prime})
-\int_{\Omega}d^{\rm Dim}x{\cal L}(x)\nonumber\\
&=&\int{d^{\rm Dim}x^\prime}h_{\Omega^\prime}(x^\prime){\cal
 {L^\prime}}({x^\prime})
-\int{d^{\rm Dim}x}h_\Omega(x){\cal L}(x),
\label{def_action.eq}
\end{eqnarray}
where we characterize this variation by $\zeta$, 
because this variation is derived from the infinitesimal transformations
Eqs. (\ref{eq:TFx}) and (\ref{eq:TFf}). 
Since the integration variable $x^\prime$ can be replaced by $x$,
Eq.~(\ref{def_action.eq}) is
\begin{eqnarray}
 \delta_\zeta I&=&\int{d^{\rm Dim}x}h_{\Omega^\prime}(x){\cal
  {L^\prime}}(x)-\int{d^{\rm Dim}x}h_\Omega(x){\cal L}(x)\nonumber\\
&=&\int{d^{\rm Dim}x}\left(h_{\Omega^\prime}(x)-h_\Omega(x)\right){\cal
 {L^\prime}}(x)+\int{d^{\rm Dim}x}h_\Omega(x)\left[{\cal L}^\prime(x)
-{\cal L}(x)\right]\nonumber\\
&=&\int d^{\rm Dim}xh_{\delta\Omega}(x){\cal L}^\prime(x)
+\int d^{\rm Dim}xh_\Omega(x)\left[{\cal L}^\prime(x)
-{\cal L}(x)\right],
\end{eqnarray}
where $\delta\Omega:=\Omega^\prime-\Omega$ and
$h_{\delta\Omega}=-(\partial_\mu h_\Omega)
\delta_\zeta x^\mu+{\cal O}((\delta_\zeta x)^2)$. 
Therefore, we obtain the following equation through partial integration:
\begin{eqnarray}
 \delta_\zeta I&=&\int{d^{\rm Dim}x}h_\Omega(x)
\left[\partial_\mu({\cal L}(x)\delta_\zeta x^\mu)
 +\delta^{\rm L}_\zeta{\cal L}(x)\right]
+{\cal O}((\delta_\zeta x)^2),
 \label{eq:VariationAction}
\end{eqnarray}
where we have introduced the so-called Lie derivative: 
\begin{eqnarray}
 \delta^{\rm L}_\zeta\phi_r(x)&:=&\phi^\prime_r(x)-\phi_r(x)\nonumber\\
 &=&\delta_\zeta\phi_r(x)-(\partial_\mu\phi_r)\delta_\zeta x^\mu+{\cal
 O}(\delta_\zeta x^2),
\end{eqnarray}
and we replaced ${\cal L}^\prime$ by ${\cal L}$ due to
${\cal L}^\prime\delta_\zeta x^\mu=
{\cal L}\delta_\zeta x^\mu+{\cal O}(\delta_\zeta x^2)$. 

Hereafter, we assume that the action is invariant under 
the infinitesimal transformations Eqs.~(\ref{eq:TFx}) and (\ref{eq:TFf}). 
In the case where the Lagrangian density is a function of $\phi_r$ and
$\partial_\mu\phi_r$, i.e., 
${\cal L}(x)={\cal L}\left[\phi_r(x),\partial_\mu\phi_r(x)\right]$, 
the Lie derivative of the Lagrangian is given by
\begin{eqnarray}
 \delta^{\rm L}_\zeta{\cal L}&:=&
{\cal L}^\prime(x)-{\cal L}(x)\nonumber\\
&=&{\cal L}\left[\phi_r^\prime(x),\partial_\mu\phi_r^\prime(x)\right]-
{\cal L}\left[\phi_r(x),\partial_\mu\phi_r(x)\right]\nonumber\\
&=&\left\{
\frac{\partial {\cal L}}{\partial\phi_r}\delta^{\rm L}_\zeta\phi_r+
\frac{\partial {\cal L}}{\partial\left(\partial_\mu\phi_r\right)}
\partial_\mu\delta^{\rm L}_\zeta\phi_r\right\}\nonumber\\
&=&\left(\frac{\partial{\cal L}}{\partial\phi_r}
-\partial_\mu\frac{\partial
{\cal L}}{\partial(\partial_\mu\phi_r)}\right)
\delta^{\rm L}_\zeta\phi_r
+\partial_\mu
\left(\frac{\partial{\cal L}}{\partial(\partial_\mu\phi_r)}
\delta^{\rm L}_\zeta\phi_r\right),
\label{lie.eq}
\end{eqnarray}
and the variation of the action is calculated by
\begin{equation}
 \delta_\zeta I=\int_\Omega d^{\rm Dim}x\left\{
\left(\frac{\partial{\cal L}}{\partial\phi_r}
-\partial_\mu
\frac{\partial{\cal L}}{\partial(\partial_\mu\phi_r)}\right)
\delta^{\rm L}_\zeta\phi_r+
\partial_\mu\left(
{\cal L}\delta_\zeta x^\mu+
\frac{\partial{\cal L}}{\partial(\partial_\mu\phi_r)}
\delta^{\rm L}_\zeta\phi_r\right)\right\}.\label{eq:VariationAction2}
\end{equation}
If we require that $\delta_\zeta x$ and $\delta^{\rm L}_\zeta\phi_r$
vanish on the surface $\partial\Omega$, 
we obtain the Euler-Lagrange equation. 
On the other hand, 
if we require that fields $\phi_r$ satisfy the Euler-Lagrange equation, 
we obtain continuity equation $\partial_\mu j^\mu=0$ with a Noether current 
\begin{equation}
 j^\mu:=\left(
{\cal L}\delta _\zeta x^\mu+
\frac{\partial{\cal L}}{\partial(\partial_\mu\phi_r)}
\delta^{\rm L}_\zeta\phi_r\right).
\end{equation}

Hereafter let us discuss an infinitesimal global U(1)$\times$SU(2) gauge
transformation and infinitesimal translation and rotation
transformations, 
which are denoted
by $\chi$ in this paper. 
Variations in terms of $\chi$ are defined by
\begin{eqnarray}
 \delta_\chi x^\mu&:=&\Gamma^\mu_\nu x^\nu,\\
 \delta_\chi\phi_r&:=&
\mbox{i}\vartheta^{\mu\nu}(\xi_{\mu\nu})_r^{r^\prime}\phi_{r^\prime}
\end{eqnarray}
with an infinitesimal parameter $\vartheta^{\mu\nu}$, 
and symmetry generators $\Gamma^\mu_\nu$ and $(\xi_{\mu\nu})_r^{r^\prime}$. 
\begin{enumerate}
 \item For the global U(1)$\times$SU(2) gauge transformation, 
$\Gamma^\mu_\nu\equiv 0$, 
$\vartheta^{\mu\nu}\equiv\vartheta^\mu\delta^{\mu\nu}$, and 
$(\xi_{\mu\nu})_r^{r^\prime}\equiv\delta_{\mu\nu}(\hat{s}_\mu)_r^{r^\prime}$ 
($\mu,\nu=0,1,2,3$; $r,r^\prime=1,2$), 
where $\hat{s}^0:=\hbar/2$, and 
$\hat{s}^{1,2,3}:=\hbar\hat{\sigma}^{x,y,z}/2$ 
with the Planck constant $h= 2 \pi \hbar$ and
Pauli matrices $\hat{\sigma}^{x,y,z}$. 
 \item For the translation, $\Gamma^\mu_\nu\equiv
\varepsilon^\mu\delta^\mu_\nu$,
$\vartheta^{\mu\nu}\equiv\varepsilon^\mu\delta^{\mu\nu}$ 
and 
$(\xi_{\mu\nu})_r^{r^\prime}\equiv\hat{p}_\mu\delta_{\mu\nu}\delta_r^{r^\prime}$
with an infinitesimal parameter $\varepsilon^\mu$ and 
the momentum operator $\hat{p}_\mu=-{\rm i}\hbar\partial_\mu$
($\mu,\nu=1,2,3$; $r,r^\prime=1,2$). 
 \item For the rotation, 
$\Gamma^\mu_\nu\equiv\omega^\mu_\nu$,
$\vartheta^{\mu\nu}\equiv\omega^{\mu\nu}$,  
and $(\xi_{\mu\nu})_r^{r^\prime}\equiv\delta_r^{r^\prime}x_\mu\hat{p}_\nu$, 
which corresponds to the angular momentum tensor 
($\mu,\nu=1,2,3$; $r,r^\prime=1,2$). 
\end{enumerate}

For these transformations, 
equation $\partial_\mu\delta_\chi x^\mu=0$ is satisfied. 
This can be seen explicitly as follows. 
The variations of space coordinates of the global U(1)$\times$SU(2) 
and the translation transformations are given by 
$\delta_\chi x^\mu=0$ or $\delta_\chi x^\mu={\rm constant}$, respectively,  
and thus $\partial_\mu\delta_\chi x^\mu=0$ is trivial.  
The variation of the rotation transformation is given by
$\delta_\chi x^\mu=\omega^\mu_\nu x^\nu$, therefore 
$\partial_\mu\delta_\chi x^\mu=
\partial_\mu\omega^\mu_\nu x^\nu=\omega^\mu_\mu=0$. 

We consider a variation of the Lagrangian density;
\begin{eqnarray}
 \delta_\zeta{\cal L}&:=&{\cal L}^\prime(x^\prime)-{\cal L}(x)\nonumber\\
&=&{\cal L}^\prime(x^\prime)-{\cal L}(x^\prime)+{\cal L}(x^\prime)-
{\cal L}(x)\nonumber\\ 
&=&\delta^{\rm L}_\zeta{\cal L}(x^\prime)+
\delta_\zeta x^\mu\partial_\mu{\cal L}+{\cal O}((\delta_\zeta x)^2).\label{eq:VL}
\end{eqnarray}
Note that Eq.~(\ref{eq:VL}) is correct 
for any infinitesimal transformation. 
Here we consider the global U(1)$\times$SU(2) gauge transformation and/or
the translation and rotation transformations $\delta_\chi$. 
Because $\partial_\mu\delta_\chi x^\mu=0$, we obtain the following equation:
\begin{equation}
 \delta_\chi{\cal L}=\delta^{\rm L}_\chi{\cal L}(x)+
\partial_\mu({\cal L}(x)\delta_\chi x^\mu)+{\cal
 O}((\delta x)^2).\label{eq:VariationLagrangian}
\end{equation}
From Eqs.(\ref{eq:VariationAction}) and (\ref{eq:VariationLagrangian}), 
one can see 
\begin{equation}
 \delta_{({\zeta=\chi})} I=\int d^{\rm Dim}x\delta_\chi{\cal L}(x),
\end{equation}
where ${(\zeta=\chi)}$ 
denotes that the type of the variation in Eq. (\ref{eq:VariationAction})
is restricted to the global U(1)$\times$SU(2) or Poincare
transformations. 
(For simplicity we omitted the subscript $\Omega$ in the integral). 
Finally, for $\zeta=\chi$, the variation of the action is equal to the
variation of Lagrangian. 
This fact will be used later in section \ref{section:TS} where the 
variation of the Lagrangian density instead of the action will be
considered.

\subsection{Generalization of Noether's theorem}\label{subsec:general}
Now, we would like to introduce a Hopf algebra for the purpose
of generalizing Noether's theorem~
\cite{Gonera2005192,PTPS.171.65,AmelinoCamelia2009298}. 
At first, we rewrite Noether's theorem in section
\ref{sec:Noether} by using the Hopf algebra, and next, 
we introduce a twisted symmetry~
\cite{Chaichian200498,PhysRevLett.94.151602}. 
For simplicity, we only consider the global U(1)$\times$SU(2) gauge
symmetry and the Poincare symmetry. 
We assume that the Lagrangian
density is written as
\begin{equation}
 {\cal L}(x)=\psi^\dagger(x)\hat{{\cal L}}(x)\psi(x)
\end{equation}
with a field 
$\psi:=\left(
\begin{array}{c}
\psi_1\\
\psi_2
\end{array}
\right)
$, a Hermitian conjugate
$\psi^\dagger\equiv\left(\overline{\psi}_1,\overline{\psi}_2\right)$, 
and an single-particle Lagrangian density operator $\hat{\cal L}$, which is
a 2$\times$2 matrix; 
the overline represents the complex conjugate. 
The action can be rewritten as
\begin{eqnarray}
 I&=&\int d^{\rm Dim}x_1d^{\rm Dim}x_2\psi^\dagger(x_2)\delta(x_2-x_1)
\hat{{\cal L}}(x_1)\psi(x_1)\nonumber\\
 &=&{\rm tr}\int d^{\rm Dim}x_1d^{\rm Dim}x_2\delta(x_2-x_1)
\hat{{\cal L}}(x_1)\psi(x_1)\psi^\dagger(x_2)\nonumber\\
 &=&{\rm tr}\int d^{\rm Dim}x_1\left\{
\int d^{\rm Dim}x_2 {\tilde{\cal L}}(x_1,x_2)G(x_2,x_1)
\right\}\nonumber\\
 &=&{\rm tr}\int d^{\rm Dim}x_1\left\{
\lim_{x_3\rightarrow x_1}
(\tilde{\cal L}*_CG)(x_1,x_3)\right\},
\label{eq:action}
\end{eqnarray}
where ``${\rm tr}$'' represents the trace in the spin space,
$G(x_1,x_2):=\psi(x_1)\psi^\dagger(x_2)$, 
$\tilde{{\cal L}}(x_1,x_2):=\delta(x_1-x_2)\hat{{\cal L}}(x_2)$, 
and $*_C$ represents
the convolution integral: 
\begin{equation}
 (f*_Cg):=\int d^{\rm Dim}x_3f(x_1,x_3)g(x_3,x_2)
\end{equation}
with smooth two-variable functions $f$ and $g$.

The variation operator $\delta_\chi$ of the action can be also rewritten as
\begin{eqnarray}
 \delta_\chi I
 &=&{\rm tr}\int d^{\rm Dim}x_1d^{\rm Dim}x_2
 \tilde{\cal L}(x_2,x_1)\left[
i\vartheta\xi\psi(x_1)\psi^\dagger(x_2)
-\psi(x_1)\psi^\dagger(x_2)i\vartheta\xi\right]\nonumber\\
 &=&{\rm tr}\int d^{\rm Dim}x_1d^{\rm Dim}x_2\left[
\tilde{{\cal L}}(x_2,x_1)
i\vartheta\xi G(x_1,x_2)-i\vartheta\xi
\tilde{{\cal L}}(x_2,x_1)G(x_1,x_2)\right]\nonumber\\ 
\label{eq:VariationXi}
\end{eqnarray}
with $\vartheta\xi\equiv\vartheta^{\mu\nu}(\xi_{\mu\nu})$; 
in addition, we assumed that the single-particle Lagrangian density
operator is invariant under the infinitesimal transformation $\delta_\chi$. 

Here, we introduce Grassmann numbers $\theta_1$ and $\theta_2$; 
an integral is defined by 
$\int d\theta_i\left(\theta_j\right)=\delta_{ij}$. 
The variation of the right-hand side of Eq.(\ref{eq:VariationXi}) 
can be rewritten as follow: 
\begin{eqnarray}
 \delta_\chi I
 &=&-{\rm i}{\rm tr}\int d\theta_1d\theta_2d^{\rm Dim}x_1d^{\rm Dim}x_2
\left[\theta_1\tilde{\cal L}(x_2,x_1)\vartheta\theta_2\xi G(x_1,x_2)
+\vartheta\theta_2\xi\theta_1\tilde{\cal L}(x_2,x_1)G(x_1,x_2)
\right]\nonumber\\
 &=&-{\rm i}{\rm tr}\int d\theta_1d\theta_2d^{\rm Dim}x_1d^{\rm Dim}x_2
\mu\circ(\mu\otimes\mbox{id})\nonumber\\
&&\circ \left[\left(
\theta_1\tilde{\cal L}(x_2,x_1)\otimes\vartheta\theta_2\xi
+\vartheta\theta_2\xi\otimes\theta_1\tilde{\cal L}(x_2,x_1)
\right)\otimes G(x_1,x_2)\right]\nonumber\\
 &=&-{\rm i}{\rm tr}\int d\theta_1d\theta_2d^{\rm Dim}x_1d^{\rm Dim}x_2
\mu\circ(\mu\otimes {\rm id})\left[\triangle(\vartheta\theta_2\xi)\circ
\left(\theta_1\tilde{\cal L}(x_2,x_1)\otimes G(x_1,x_2)\right)\right]\nonumber\\
 &\equiv&{\rm \hat{T}r}\left[\triangle(\vartheta\theta_2\xi)\circ
((\theta_1\tilde{\cal L})\otimes G)\right],\label{eq:VariableOfActionByCoproduct}
\end{eqnarray}
where $\otimes$ and $\circ$ represent a tensor product and a product of
operators, respectively. 
The operator $\mu$ denotes the transformation of 
the tensor product to the usual product 
$\mu:x\otimes y\mapsto xy$, 
and $\triangle$ represents a coproduct:
\begin{equation}
 \triangle(\zeta):=\zeta\otimes {\rm id}+{\rm id}\otimes\zeta,
\label{eq:coproductz}
\end{equation}
where $\zeta$ and ${\rm id}$ represent a certain operator and the identity map,
respectively. 
These operators constitutes the Hopf algebra as will be
explained in the next section. 
Moreover, we have defined 
${\rm \hat{T}r}:=-{\rm i}{\rm tr}\int 
d\theta_1d\theta_2d^{\rm Dim}x_1d^{\rm Dim}x_2\mu\circ(\mu\otimes {\rm id})$. 
We emphasize here that the variation is written by the coproduct
$\triangle$, 
which is important to formulate the generalized Noether theorem in the 
presence of the gauge potential. 
The coproduct determines an operation rule of a variation operator; 
for example, the coproduct (\ref{eq:coproductz}) represents 
the Leibniz rule. 
A twisted symmetry transformation is given by deformation of the coproduct.

We now sketch the concept of the twisted symmetry
in deformation quantization~\cite{Chaichian200498,PhysRevLett.94.151602}. 
First, we assume that the variation of action $\delta_\zeta I_0$ is
zero, i.e., $\zeta$ represents the symmetry transformation of the 
system corresponding to the action $I_0$. 
Next, we consider the action $I_A$ with external gauge fields $A$. 
Usually, external gauge fields breaks symmetries of $I_0$, i.e.,
$\delta_\zeta I_A\neq 0$. 
Here we introduce a map: ${\cal F}_{(0\mapsto A)}:I_0\mapsto I_A$,
which will be defined in section \ref{sec:TwistedElement}.
The basic idea is to generalize the "product" taking into account
the gauge interaction.
Using this map, the variation is rewritten as 
$\delta_\zeta {\cal F}_{(0\mapsto A)}I_0\neq 0$. 
On the other hand, when the twisted symmetry 
$\delta^t_\zeta:={\cal F}_{(0\mapsto A)}\delta_\zeta{\cal
F}^{-1}_{(0\mapsto A)}$ 
can be defined, we obtain the following equation: 
\begin{eqnarray} 
 \delta^t_\chi I_A&=&{\cal F}_{(0\mapsto A)}\delta_\chi 
{\cal F}^{-1}_{(0\mapsto A)}I_A\nonumber\\
&=&{\cal F}_{(0\mapsto A)}\delta_\chi {\cal F}^{-1}_{(0\mapsto A)}
{\cal F}_{(0\mapsto A)}I_0\nonumber\\ 
&=&{\cal F}_{(0\mapsto A)}\delta_\chi I_0\nonumber\\
&=&0.
\end{eqnarray}
Namely, $\delta^t_\chi$ corresponds to a symmetry with external gauge
fields. 
In the expression for the variation of action in terms of the Hopf algebra 
Eq.(\ref{eq:VariableOfActionByCoproduct}), 
we can replace $\Delta$ by $\Delta^t$ corresponding to the change from 
$\delta\chi$ to $\delta^t_\chi$ as shown in section \ref{section:TS}.
This is achieved by using the Hopf algebra and the deformation
quantization, which will be explained in sections 
\ref{sec:Hopf} and \ref{sec:star}, respectively.
Therefore, we can generalize the Noether's theorem 
and derive the conservation law even in the presence
of the gauge field $A$.

\section{Hopf algebra}\label{sec:Hopf}

Here we introduce a Hopf algebra. 
First, we rewrite the algebra using tensor and linear maps. 
Secondly, a coalgebra is defined using diagrams corresponding to the algebra. 
Finally, we define a dual-algebra and Hopf algebra. 

\subsubsection{Algebra}\label{subsec:Alge}
We define the algebra as a $k$-vector space $V$ having product $\mu$ and
unit $\varepsilon$. 
Here, $k$ represents a field such as the complex number or real number. 
In this paper, we consider $V$ as the space of functions or operators. 
A space of linear maps from a vector space $V_1$ to a vector space
$V_2$ is written as ${\rm Hom}(V_1,V_2)$. 

A product $\mu$ is a bilinear map: $\mu\in {\rm Hom}(V\bigotimes V,V)$, i.e.,
\begin{equation}
 \mu: V\bigotimes V\rightarrow V,\ \ \ \ \ (x,y)\mapsto xy,
\end{equation}
and a unit is a linear map: $\varepsilon\in {\rm Hom}(k,V)$, i.e., 
\begin{equation}
 \varepsilon:k\rightarrow V,\ \ \ \ \ \alpha\mapsto \alpha\cdot 1
\end{equation}
with $x,y,xy\in V$ and $\alpha\in k$. 
Here $\mu$ and $\varepsilon$ satisfies
\begin{equation}
 \mu((x+y)\otimes z)=\mu(x\otimes z)+
\mu(y\otimes z),~\mu(x\otimes(y+z))=
\mu(x\otimes y)+\mu(x\otimes z),
\end{equation}
\begin{equation}
 \mu(\alpha x\otimes y)=
\alpha\mu(x\otimes y),~\mu(x\otimes \alpha y)=
\alpha\mu(x\otimes y),
\end{equation}
\begin{equation}
 \varepsilon(\alpha+\beta)=\varepsilon(\alpha)+\varepsilon(\beta)
\end{equation}
with $x,y,z\in V$ and $\alpha,\beta\in k$. 

The product $\mu$ has the association property, which is written as
$\mu\circ({\rm id}\otimes\mu)=\mu\circ(\mu\otimes {\rm id})$. 
Because the left-hand side and the right-hand side of 
the previous equation give the following equations:
\begin{equation}
 \mu\circ({\rm id}\otimes\mu)(x\otimes y\otimes z)=\mu(x\otimes (yz))=x(yz)
\end{equation}
and 
\begin{equation}
 \mu\circ(\mu\otimes {\rm id})(x\otimes y\otimes z)=\mu((xy\otimes z))=(xy)z,
\end{equation}
for all $x,y,z,xy,yz,xyz\in V$,
then $\mu\circ({\rm id}\otimes \mu)=\mu\circ(\mu\otimes {\rm id})$ is equal to
the association property $x(yz)=(xy)z$. 
This property is illustrated as the following diagram:
\vspace{1.5cm}
$$
\begin{picture}(90,50)(0,-20)
 \put(-18,50){$V\bigotimes V\bigotimes V$}
 \put(100,50){$V\bigotimes V$}
 \put(-10,0){$V\bigotimes V$}
 \put(107,0){$V$}
 \put(6,45){\vector(0,-1){30}}
 \put(40,54){\vector(1,0){50}}
 \put(40,3){\vector(1,0){50}}
 \put(113,45){\vector(0,-1){30}}
 \put(50,58){$\mu\otimes {\rm id}$}
 \put(-24,30){${\rm id}\otimes \mu$}
 \put(117,30){$\mu$}
 \put(60,8){$\mu$}
 \put(65,30){\cbox{$\circlearrowright$}}
\end{picture}
$$
Here $\circlearrowright$ denotes that this graph is the commutative
diagram. 

The unit $\varepsilon$ satisfies the following equation:
$\mu\circ(\varepsilon\otimes {\rm id})=\mu\circ({\rm id}\otimes\varepsilon)$.
Since the left-hand side and the right-hand side of the previous
equation give the following equations
\begin{equation}
 \mu\circ(\varepsilon\otimes {\rm id})(\alpha\otimes x)=\mu\otimes(\alpha
  1_V\otimes x)=\alpha 1_Vx=\alpha x
\end{equation}
and 
\begin{equation}
 \mu\circ({\rm id}\otimes\varepsilon)(x\otimes
  \alpha)=\mu\circ(x\otimes \alpha 1_V)=\alpha x1_V=\alpha x
\end{equation}
for all $x\in V$ and $\alpha\in k$, 
and ${}^\exists 1_V\in V$, 
then the unit can be written as
$\mu\circ({\rm id}\otimes\varepsilon)=\mu\circ(\varepsilon\otimes {\rm id})$.
Note that $V\sim k\bigotimes V\sim V\bigotimes k$, where $\sim$
represents the equivalence relation, i.e., $a\sim b$
denotes that $a$ and $b$ are identified. 
This property is illustrated as: 
$$
\begin{picture}(180,70)(0,0)
 \put(0,50){$k\bigotimes V$}
 \put(180,50){$V\bigotimes k$}
 \put(90,50){$V\bigotimes V$}
 \put(102,0){$V$}
 \put(40,52){\vector(1,0){45}}
 \put(105,43){\vector(0,-1){30}}
 \put(170,52){\vector(-1,0){45}}
 \put(20,43){\vector(2,-1){70}}
 \put(190,43){\vector(-2,-1){70}}
 \put(50,55){$\varepsilon\otimes {\rm id}$}
 \put(135,55){${\rm id}\otimes\varepsilon$}
 \put(95,30){$\mu$}
 \put(40,20){$\sim$}
 \put(165,20){$\sim$}
\end{picture}
$$
Algebra is defined as a set $(V,\mu,\varepsilon)$. 

\subsubsection{Coalgebra}\label{subsubsec:Co}
A coalgebra is defined by reversing the direction of the arrows in the diagrams
corresponding to the algebra. 
Thus, we will define a coproduct $\triangle\in {\rm Hom}(V,V\bigotimes V)$ and
counit $\eta\in {\rm Hom}(V,V)$ with a $k$-vector space $V$. 

A coproduct is a bilinear map from $V$ to $V\bigotimes V$:
\begin{equation}
 \triangle:V\rightarrow V\bigotimes V,
\end{equation}
and satisfies co-association property:
\vspace{1.5cm}

$$
\begin{picture}(90,50)(0,-20)
 \put(-18,50){$V\bigotimes V\bigotimes V$}
 \put(100,50){$V\bigotimes V$}
 \put(-10,0){$V\bigotimes V$}
 \put(107,0){$V$}
 \put(6,12){\vector(0,1){30}}
 \put(90,54){\vector(-1,0){50}}
 \put(90,3){\vector(-1,0){50}}
 \put(113,12){\vector(0,1){30}}
 \put(50,58){$\triangle\otimes {\rm id}$}
 \put(-28,30){${\rm id}\otimes \triangle$}
 \put(117,30){$\triangle$}
 \put(60,8){$\triangle$}
 \put(65,30){\cbox{$\circlearrowright$}}
\end{picture}
$$

Namely,
\begin{equation}
 ({\rm id}\otimes\triangle)\circ\triangle=(\triangle\otimes {\rm id})\circ\triangle
\end{equation}
(Compare the diagram corresponding to the association property and that
corresponding to the co-association property).

A counit $\eta$ is a linear map from $V$ to field $k$:
\begin{equation}
 \eta: V\rightarrow k,
\end{equation} 
and satisfies the following diagram:

$$
\begin{picture}(180,70)(0,0)
 \put(0,50){$k\bigotimes V$}
 \put(180,50){$V\bigotimes k$}
 \put(90,50){$V\bigotimes V$}
 \put(102,0){$V$}
 \put(85,52){\vector(-1,0){45}}
 \put(105,13){\vector(0,1){30}}
 \put(125,52){\vector(1,0){45}}
 \put(90,8){\vector(-2,1){70}}
 \put(117,8){\vector(2,1){70}}
 \put(50,55){$\eta\otimes {\rm id}$}
 \put(135,55){${\rm id}\otimes\eta$}
 \put(95,25){$\triangle$}
 \put(40,20){$\sim$}
 \put(165,20){$\sim$}
\end{picture}
$$

Namely, 
\begin{equation}
 (\eta\otimes {\rm id})\circ\triangle=({\rm id}\otimes\eta)\circ\triangle,
\end{equation}
where $V\sim k\bigotimes V\sim V\bigotimes k$. 

Since $\triangle$ and $\eta$ are linear maps, $\triangle$ and
$\eta$ satisfy
\begin{equation}
 \triangle(x+y)=\triangle(x)+\triangle(y),~\triangle(\alpha
  x)=\alpha\triangle(x), 
\end{equation}
\begin{equation}
 \eta(x+y)=\eta(x)+\eta(y),~\eta(\alpha x)=\alpha\eta(x)
\end{equation}
with $x,y\in V$ and $\alpha\in k$. 
Note that $V\sim k\bigotimes V\sim V\bigotimes k$ and $V\bigotimes V\sim
k\bigotimes V\bigotimes V\sim V\bigotimes k\bigotimes V\sim V\bigotimes
V\bigotimes k$. 

A coalgebra is defined as a set $(V,\triangle,\eta)$. 
For example, in the vector space 
$D\equiv k\bigoplus
k\partial:=\{a_0+a_1\partial|a_0,a_1\in k\}$, 
we define a coproduct
$\triangle_D(\partial)=\partial\otimes 1+1\otimes\partial$ and
$\triangle_D(1)=1\otimes 1$, and a counit $\eta_D(\partial)=0$ and
$\eta_D(1)=1$. 
The set $(D,\triangle_D,\eta_D)$ is coalgebra, because this set
satisfies the equations: $(\triangle_D\otimes {\rm id})\circ
\triangle_D=({\rm id}\otimes\triangle_D)\circ\triangle_D$ and 
$(\eta_D\otimes {\rm id})\circ\triangle_D=({\rm id}\otimes\eta_D)\circ\triangle_D$. 
Because the coproduct and counit are linear map, we only check the above
equations with respect to $x=1$ and $\partial$. 

For $x=1$, 
\begin{equation}
 (\triangle_D\otimes {\rm id})\circ\triangle_D(1)=\triangle_D(1)\otimes 1
=1\otimes 1\otimes 1,
\end{equation}
and
\begin{equation}
 ({\rm id}\otimes\triangle_D)\circ\triangle_D(1)=1\otimes\triangle_D(1)=1\otimes
  1\otimes 1.
\end{equation}
Therefore, $(\triangle_D\otimes
{\rm id})\circ\triangle_D(1)=({\rm id}\otimes\triangle_D)\circ\triangle_D(1)$. 
Moreover, 
\begin{equation}
 (\eta_D\otimes {\rm id})\circ\triangle_D(1)=\eta_D(1)\otimes
  1=1\otimes 1,
\end{equation}
and
\begin{equation}
 ({\rm id}\otimes\eta_D)\circ\triangle_D(1)=1\otimes\eta_D(1)=1\otimes 1.
\end{equation}
Therefore $(\eta_D\otimes
{\rm id})\circ\triangle_D(1)=({\rm id}\otimes\eta_D)\circ\triangle_D(1)$. 

For $x=\partial$, 
\begin{equation}
 (\triangle_D\otimes {\rm id})\circ\triangle_D(\partial)=
\triangle_D(\partial)\otimes 1+\triangle_D(1)\otimes \partial=
\partial\otimes 1\otimes 1+1\otimes \partial\otimes 1+1\otimes 1\otimes\partial,
\end{equation}
and
\begin{equation}
 ({\rm id}\otimes\triangle_D)\circ\triangle_D(\partial)=
\partial\otimes\triangle_D(1)+1\otimes\triangle_D(\partial)=
\partial\otimes 1\otimes 1+1\otimes \partial\otimes 1+1\otimes 1\otimes\partial.
\end{equation}
Therefore,
$({\rm id}\otimes\triangle_D)\circ\triangle_D(\partial)=(\triangle_D\otimes
{\rm id})\circ\triangle_D(\partial)$. 
Finally, 
\begin{equation}
 (\eta_D\otimes {\rm id})\circ\triangle_D(\partial)=
\eta_D(\partial)\otimes 1+\eta_D(1)\otimes\partial=
1\otimes\partial=
\partial,
\end{equation}
and 
\begin{equation}
 ({\rm id}\otimes\eta_D)\circ\triangle_D(\partial)=
\partial\otimes\eta_D(1)+1\otimes\eta_D(\partial)=
\partial\otimes 1=
\partial.
\end{equation}
Therefore, 
$(\eta_D\otimes {\rm id})\circ
\triangle_D(\partial)=
({\rm id}\otimes\eta_D)\circ
\triangle_D(\partial)$. 
Namely, the set $(D,\triangle_D,\eta_D)$ is the coalgebra. 
Note that $\triangle_D(1)$ corresponds to the product with a
constant: $a(fg)=a1(fg)=a(1f1g)=a\mu\circ\triangle_D(1)(f\otimes g)$, 
where we have used the coproduct $\triangle_D(1)=1\otimes 1$ at the
final equal sign. 
Here $f,~{\rm and}~g$ are smooth functions, $1$ is included in the
function space, and $a\in k$. 
$\triangle_D(\partial)$ represents the Leibniz rule: 
$\partial(fg)=(\partial
f)g+f\partial(g)=\mu\circ(\partial\otimes
1+1\otimes\partial)\circ(f\otimes
g)=\mu\circ\triangle_D(\partial)(f\otimes g)$, 
where we have used the coproduct 
$\triangle_D(\partial)=1\otimes\partial+\partial\otimes 1$ at
the last equal sign. 
$\eta_D(1)$ and $\eta_D(\partial)$ represent the filtering action to a constant function:
$1a=a=\eta_D(1)a$ and $\partial(a)=0=\eta_D(\partial)a$, respectively. 

\subsubsection{Dual-algebra and Hopf algebra}\label{subsec:DH}
A dual-algebra is the set of an algebra and a coalgebra, i.e., the set of
$(V,\mu,\varepsilon,\triangle,\eta)$. 
On a dual-algebra, we define a $*$-product as
\begin{equation}
 f*g=\mu\circ(f\otimes g)\circ \triangle
\end{equation}
with $f,g\in{\rm Hom}(V,V)$. 
We define an antipode $S\in {\rm Hom}(V,V)$ which satisfies the following
equation: 
\begin{equation}
 \mu\circ({\rm id}\otimes S)\circ \triangle=\mu\circ(S\otimes {\rm id})\circ
  \triangle=\varepsilon\circ\eta, 
\end{equation}
where $\varepsilon\circ\eta$ corresponds to the identity mapping, i.e.,
$S$ is an inverse of unit. 
For example, $S_D$ in the set
$(D,\mu_D,\varepsilon_D,\triangle_D,\eta_D)$ is defined as
$S_D(1)=1$ and $S_D(\partial)=-\partial$. 

For $x=1$, 
\begin{equation}
 \mu_D\circ({\rm id}\otimes S_D)\circ\triangle_D(1)=
\mu_D\circ(1\otimes 1)=1,
\end{equation}
and
\begin{equation}
 \mu_D\circ(S_D\otimes {\rm id})\circ\triangle_D(1)=
\mu_D\circ(1\otimes 1)=1.
\end{equation}
Therefore, we obtain 
$\mu_D\circ({\rm id}\otimes S_D)\circ
\triangle_D(1)=\mu_D\circ(S_D\otimes {\rm id})
\circ\triangle_D(1)=\varepsilon_D\circ\eta_D$. 
For $\partial$, 
\begin{equation}
 \mu_D\circ({\rm id}\otimes S_D)\circ
 \triangle_D(\partial)=
 \mu_D\circ(\partial\otimes 1-
 1\otimes \partial)=0,
\end{equation}
and
\begin{equation}
 \mu_D\circ(S_D\otimes {\rm id})\circ\triangle_D(\partial)=
 \mu_D\circ(-\partial\otimes 1+1\otimes \partial)=0.
\end{equation}
Therefore, we obtain 
$\mu_D\circ({\rm id}\otimes
S_D)\circ\triangle_D(\partial)=
\mu_D\circ(S_D\otimes
{\rm id})\circ\triangle_D(\partial)=\varepsilon_D\circ\eta_D(\partial)$. 
Namely, 
$(D,\mu_D,\varepsilon_D,\triangle_D,\eta_D)$ 
is the Hopf algebra. 

A dual-algebra with an antipode $S$, i.e.,
$(V,\mu,\varepsilon,\triangle,\eta,S)$, is called a Hopf algebra. 

By using the approach similar to a coproduct and counit, we can define
a codifferential operator $Q\in{\rm Hom(V,V)}$ 
from a diagram of the differential $\partial\in{\rm Hom}(V,V)$. 
The differential $\partial$ is the linear map: 
\begin{equation}
 \partial:V\rightarrow V,
\end{equation}
and satisfies Leibniz rule
\begin{equation}
 \partial\circ\mu=\mu\circ({\rm id}\otimes\partial+\partial\otimes {\rm id}),
\end{equation}
which is illustrated as
\vspace{1.5cm}
$$
\begin{picture}(90,50)(0,-20)
 \put(-20,50){$V$}
 \put(110,50){$V$}
 \put(-30,0){$V\bigotimes V$}
 \put(100,0){$V\bigotimes V$}
 \put(-16,12){\vector(0,1){30}}
 \put(90,54){\vector(-1,0){80}}
 \put(90,3){\vector(-1,0){80}}
 \put(113,12){\vector(0,1){30}}
 \put(50,58){$\partial$}
 \put(-28,27){$\mu$}
 \put(117,27){$\mu$}
 \put(12,8){$({\rm id}\otimes\partial+\partial\otimes {\rm id})$}
 \put(48,30){\cbox{$\circlearrowright$}}
\end{picture}
$$
A codifferential operator $Q$ is a linear map; $Q:V\rightarrow V$, and satisfies
the following diagrams: 
\vspace{1.5cm}
$$
\begin{picture}(90,50)(0,-20)
 \put(-20,50){$V$}
 \put(110,50){$V$}
 \put(-30,0){$V\bigotimes V$}
 \put(100,0){$V\bigotimes V$}
 \put(-16,42){\vector(0,-1){30}}
 \put(10,54){\vector(1,0){80}}
 \put(10,3){\vector(1,0){80}}
 \put(113,42){\vector(0,-1){30}}
 \put(50,58){$Q$}
 \put(-28,27){$\triangle$}
 \put(117,27){$\triangle$}
 \put(12,8){$({\rm id}\otimes Q+Q\otimes {\rm id})$}
 \put(48,30){\cbox{$\circlearrowright$}}
\end{picture}
$$
Namely, a codifferential operator $Q$ satisfies 
$\triangle\circ Q=({\rm id}\otimes Q+Q\otimes {\rm id})\circ\triangle$. 
In section \ref{sub:L_infty}, 
the codifferential operator will be introduced. 

\section{Deformation quantization}\label{sec:star}

In this section, we explain the deformation quantization 
using the noncommutative product encoding the 
commutation relationships.
At first, in section \ref{sec:WR}, 
we introduce the so-called Wigner representation and 
Wigner space, and show that a product in the Wigner space 
is noncommutative. 
This product is called Moyal product and it guarantees the commutation
relationship of the coordinate and canonical momentum. 
Next, we add spin functions and background gauge fields 
to the Wigner space, 
and rewrite the coordinates of Wigner space as 
a set of spacetime coordinates $X$, mechanical momenta $p$,
and spins $s:=(s^x,s^y,s^z)$ 
($p$ includes the background gauge fields). 
To generalize the Moyal product for the deformed Wigner space, 
which is a set of function defined on $(X,p,s)$, 
we explain the general constructing method of the noncommutative
product in section \ref{sec:SP}; 
the noncommutative product is the generalized Moyal product, 
which is called ``star product''. 
This constructing method is given as a map from 
a Poisson bracket in the Wigner space
to the noncommutative product (see section \ref{sec:SP}), 
and we see the condition of this deformation quantization map 
in section \ref{sec:SP}. 
This map is described by 
the path integral of a two-dimensional field theory,
which is called the topological string theory. 
In section \ref{sec:TopologicalStringTheory}, 
we explain this topological string theory, and in section
\ref{sec:EDQTST}, 
we discuss the perturbative treatment of this theory. 
In section \ref{sec:DiagramRules}, we summarize the diagram technique. 
Finally, in section \ref{sec:TwistedElement}, 
we construct the star product in $(X,p,s)$ space. 
We note that the star product guarantees the background gauge structure.

\subsection{Wigner representation}\label{sec:WR}

We start with the introduction of the Wigner representation.
From Equation~(\ref{eq:action}), 
a natural product is the convolution integral:
\begin{equation}
(f*_Cg)(x_1,x_2):=\int d^{\rm Dim}x_3f(x_1,x_3)g(x_3,x_2), 
\end{equation}
where $f,g\in{\cal G}$ with a two spacetime arguments function space ${\cal G}$. 
Here we introduce the center of mass coordinate $X$ and the relative
coordinate $\xi$ as follows:
\begin{eqnarray}
 X\equiv(T,\bm{X})&:=&((t_1+t_2)/2,(\bm{x}_1+\bm{x}_2)/2),\\
 \xi\equiv(\xi_t,\bm{\xi})&:=&(t_1-t_2,\bm{x}_1-\bm{x}_2).
\end{eqnarray}
Moreover we employ the following Fourier transformation:
\begin{equation}
 {\cal F}_{\cal T}:f(x_1,x_2)\mapsto f(X,p)=\int d^{\rm Dim}\xi
e^{-ip_\mu\xi^\mu/\hbar}f(X+\xi/2,X-\xi/2).
\end{equation}
Now we define the Wigner space: 
${\cal W}:=\{{\cal F}_{\cal T}[f]~|~f\in{\cal G}\}$~\cite{Wigner}.
In this space, the convolution is transformed to the so-called Moyal
product~\cite{Moyal,SOnoda}:
\begin{equation}
 (f\star _{\cal M}g)(X,p):=f(X,p)e^{\frac{{\rm i}\hbar}{2}\left({\overleftarrow
\partial}_{X}{\overrightarrow \partial}_{p_\nu}-{\overleftarrow
\partial}_{p}{\overrightarrow
\partial}_{X}\right)}g(X,p),
\end{equation}
because 
\begin{eqnarray}
{\cal F}_{\cal T}^{-1}[f\star _{\cal M}g]
&=&\int\frac{d^{\rm Dim}p}{(2\pi\hbar)^{\rm Dim}}
e^{ip_\nu\xi^\nu/\hbar}
\left\{f(X,p)
e^{\frac{{\rm i}\hbar}{2}\left({\overleftarrow
\partial}_{X^\nu}{\overrightarrow \partial}_{p_\nu}-{\overleftarrow
\partial}_{p_\nu}{\overrightarrow
\partial}_{X^\nu}\right)}g(X,p)\right\}\nonumber\\
&=&\int\frac{d^{\rm Dim}p}{(2\pi\hbar)^{\rm Dim}}d^{\rm Dim}\xi_1d^{\rm
 Dim}\xi_2e^{ip_\nu\xi^\nu/\hbar}
\Big\{e^{-ip_\nu\xi_1/\hbar}f(X+\xi_1/2,X-\xi_1/2)\nonumber\\
&&\times e^{\frac{{\rm i}\hbar}{2}\left({\overleftarrow
\partial}_{X^\nu}{\overrightarrow \partial}_{p_\nu}-{\overleftarrow
\partial}_{p_\nu}{\overrightarrow
\partial}_{X^\nu}\right)}e^{-ip_\nu\xi^\nu_2/\hbar}
g(X+\xi_2/2,X-\xi_2/2)\Big\}\nonumber\\
&=&\int\frac{d^{\rm Dim}p}{(2\pi\hbar)^{\rm
 Dim}}d^{\rm Dim}\xi_1d^{\rm Dim}\xi_2
e^{ip_\nu\left(\xi^\nu-\xi^\nu_1-\xi^\nu_2\right)/\hbar}\nonumber\\
&&\times f(X+\xi_1/2,X-\xi_1/2)e^{\frac{1}{2}
\left({\overleftarrow\partial}_{X^\nu}\xi^\nu_2-
\xi^\nu_1{\overrightarrow\partial}_{X^\nu}\right)}
g(X+\xi_2/2,X-\xi_2/2)\nonumber\\
&=&\int d^{\rm Dim}\xi_1d^{\rm Dim}\xi_2\delta(\xi-\xi_1-\xi_2)\nonumber\\
&&\times f\left(X+\frac{\xi_1+\xi_2}{2},
X-\frac{\xi_1-\xi_2}{2}\right)
g\left(X-\frac{\xi_1-\xi_2}{2},
X-\frac{\xi_1+\xi_2}{2}\right)\nonumber\\
&=&\int d^{\rm Dim}x_+d^{\rm Dim}x_-\delta(\xi-x_+)
f\left(X+\frac{x_+}{2},x_-\right)
g\left(x_-,X-\frac{x_+}{2}\right)\nonumber\\
&=&\int d^{\rm Dim}x_-f(X+\xi/2,x_-)
g(x_-,X-\xi/2)\nonumber\\
&=&f*_Cg
\end{eqnarray}
with $\xi_1+\xi_2\equiv x_+$ and 
$\xi_1-\xi_2\equiv 2(X-x_-)$. 

In the Wigner space, the position operator $\hat{x}^\mu=x^\mu$ 
and the momentum operator $\hat{p}_\mu=-{\rm i}\hbar\partial_\mu$ 
becomes $X^\mu\star_{\cal M}$ and $p_\mu\star_{\cal M}$ because
\begin{equation}
 {\cal F}_{\cal T}[\hat{x}^\mu_1g(x_1,x_2)]=
 \int d^{\rm Dim}\xi e^{-\frac{{\rm i}}{\hbar}p_\nu\xi^\nu}
 \left(X^\mu+\xi^\mu/2\right)g(X+\xi/2,X-\xi/2)
=\left(X^\mu+\frac{{\rm i}\hbar}{2}\partial_{p_\mu}\right)
g(X,p)
=X^\mu\star_{\cal M}g(X,p),
\end{equation}
\begin{equation}
 {\cal F}_{\cal T}[\hat{x}^\mu_2g(x_1,x_2)]=
 \int d^{\rm Dim}\xi e^{-\frac{{\rm i}}{\hbar}p_\nu\xi^\nu}
 g(X+\xi/2,X-\xi/2)\left(X^\mu-\xi^\mu/2\right)
=g(X,p)\star_{\cal M}X^\mu,
\end{equation}
\begin{eqnarray}
 {\cal F}_{\cal T}[(\hat{p}_1)_\mu g(x_1,x_2)]=
 \int d^{\rm Dim}\xi 
 e^{-\frac{{\rm i}}{\hbar}p_\nu\xi^\nu}
 \frac{\hbar}{{\rm i}}\partial_{x_1^\mu}g(x_1,x_2)
=p_\mu\star g(X,p),
\end{eqnarray}
and
\begin{eqnarray}
 {\cal F}_{\cal T}[(\hat{p}_2)_\mu g(x_1,x_2)]=
 \int d^{\rm Dim}\xi 
 e^{-\frac{i}{\hbar}p_\nu\xi^\nu}
 \frac{\hbar}{{\rm i}}\partial_{x_2^\mu}g(x_1,x_2)
=g(X,p)\star p_\mu.
\end{eqnarray}
The commutation relationship of operators is 
$[X^\mu,p_\nu]_{\star_{\cal M}}:=X^\mu\star_{\cal M}p_\nu
-p_\nu\star_{\cal M}X^\mu=
{\rm i}\hbar\delta^\mu_\nu$, which corresponds to the canonical
commutation relationship of operators: 
$[\hat{x}^\mu,\hat{p}_\nu]={\rm i}\hbar\delta^\mu_\nu$.

To add the spin arguments in ${\cal W}$, 
we will employ the following bilinear map: 
\begin{equation}
{\cal F}_{\cal M}\mapsto {\cal
 F}_{A=0}:=e^{\frac{{\rm i}\hbar}{2}\left(
\partial_{X^\mu}\otimes\partial_{p_\mu}-
\partial_{p_\mu}\otimes\partial_{X^\mu}\right)+
\frac{i}{2}\epsilon^{abc}s^a\partial_{s^b}\otimes
\partial_{s^c}}\label{eq:M+S}
\end{equation}
with $\partial_{s^a}f:=f_a$ and $f\equiv f_0+\sum_{a=x,y,z}s^af_a$. 
Note that the spin operator
$\bm{\hat{s}}:=(\hat{s}^x,\hat{s}^y,\hat{s}^z)$ 
is characterized by the commutation relation 
$[\hat{s}^a,\hat{s}^b]={\rm i}\epsilon^{abc}\hat{s}^c$ $(a,b,c=x,y,z)$ with 
the Levi-Civita tensor $\epsilon^{abc}$, 
and the star product (\ref{eq:M+S}) 
reproduces the relation, i.e., 
the operator $(s^a\star)$ satisfies 
$[s^a,s^b]_\star={\rm i}\epsilon^{abc}s^c$. 

To obtain the map ${\cal F}_{(0\mapsto A)}:I_0\mapsto I_A$, 
we introduce the variables transformation
$(X^\mu,p_\mu,\bm{s})\mapsto (X^\mu,\hat{p}_\mu,\bm{s})$ where
\begin{equation}
 \hat{p}_\mu=p_\mu-qA^a_\mu(X^\nu)s^a+eA_\mu
\end{equation}
with $q=|e|/{\rm m}c^2$, 
the electric charge $-e=-|e|$, a U(1) gauge field
$A_\mu$, and a SU(2) gauge field $A^a_\mu$. 
Their fields are treated as real
numbers, and the integral over $p_\mu$ can be replaced by an integral
over $\hat{p}_\mu$. 
This transformation induces the following transformations of 
differential operators:
\begin{equation}
 \partial_{X^\mu}\otimes\partial_{p_\mu}-\partial_{p_\mu}\otimes\partial_{X^\mu}
 \mapsto\partial_{X^\mu}\otimes\partial_{\hat{p}_\mu}-\partial_{\hat{p}_\mu}\otimes\partial_{X^\mu}+
 q\left(\partial_{X^\mu}\hat{A}_\nu-\partial_{X^\nu}\hat{A}_\mu\right)
 \partial_{\hat{p}_\mu}\otimes\partial_{\hat{p}_\nu},\label{eq:Transition1}
\end{equation}
\begin{eqnarray}
 \epsilon^{abc}s^a\partial_{s^b}\otimes\partial_{s^c}&\mapsto&
 \epsilon^{abc}s^a\partial_{s^b}\otimes\partial_{s^c}-
 q\epsilon^{abc}A^b_\mu s^a\partial_{\hat{p}_\mu}\otimes\partial_{s^c}-
 q\epsilon^{abc}A^c_\mu s^a\partial_{s^b}\otimes\partial_{\hat{p}_\mu}\nonumber\\
 &&+q^2\epsilon^{abc}s^aA^b_\mu A^c_\nu
 \partial_{\hat{p}_\mu}\otimes\partial_{\hat{p}_\nu},\label{eq:Transition2} 
\end{eqnarray}
where $\hat{A}_\mu:=A_\mu^as^a-(e/q)A_\mu$. 

We expand ${\cal F}_{A=0}$ in terms of $\hbar$ as
\begin{equation}
 {\cal F}_{A=0}=\sum_{n=0}^\infty\left(\frac{{\rm i}\hbar}{2}\right)^n{\cal F}_{A=0}^n.
\end{equation}
We define the bilinear map  ${\cal F}_A$ 
corresponding to the commutation relation in
terms of the phase space $(X^\mu,\hat{p}_\mu,\bm{s})$, and
expand it in terms of $\hbar$ as
\begin{equation}
 {\cal F}_A=\sum_{n=0}^\infty\left(\frac{{\rm i}\hbar}{2}\right)^n{\cal F}_A^n.
\end{equation}
From Eqs.~(\ref{eq:Transition1}) and (\ref{eq:Transition2}), 
${\cal F}_A^1$ is given as follows:
\begin{eqnarray}
 {\cal F}_A^1&=&
 \partial_{X^\mu}\otimes\partial_{\hat{p}_\mu}-\partial_{\hat{p}_\mu}
 \otimes\partial_{X^\mu}+q\hat{F}_{\mu\nu}\partial_{\hat{p}_\mu}\otimes
 \partial_{\hat{p}_\nu}+\epsilon^{abc}s^a\partial_{s^b}\otimes\partial_{s^c}\nonumber\\
 &&-q\epsilon^{abc}s^aA^b_\mu\partial_{\hat{p}_\mu}\otimes\partial_{s^c}+
 q\epsilon^{abc}s^aA^b_\mu\partial_{s^c}\otimes\partial_{\hat{p}_\mu}\label{eq:SU(2)Poisson}
\end{eqnarray}
with $\hat{F}_{\mu\nu}:=
\partial_{X^\mu}\hat{A}_\nu -\partial_{X^\nu}\hat{A}_\mu
+(q/\hbar)\varepsilon^{abc}s^a\hat{A}^b_\mu\hat{A}^b_\nu$. 
Note that $\mu\circ{\cal F}_A^1$ is the Poisson bracket. 

A constitution method of higher order terms ${\cal F}^{n}_A$ with $n>1$
is called a deformation quantization, which is given by
Kontsevich~\cite{Kontsevich}, as will be described in the next subsection.

\subsection{Star product}\label{sec:SP}
In this subsection, we explain the Kontsevich's deformation quantization
method~\cite{Kontsevich}.
We define a star product as 
\begin{equation}
 f\star g\equiv\mu\circ {\cal F}_A(f\otimes g)=
f\cdot g+\sum_{n=1}^\infty \nu^n\beta_n(f\otimes g)\label{eq:DQ0}
\end{equation}
with $\nu={\rm i}\hbar/2$~\cite{Bayen197861,Bayen1978111}. 
Here $\beta_n\in{\rm Hom}\left(V_f\otimes V_f,V_f\right)$ 
is called the two-cochain ($V_f$ represents the function space). 
We require that the star product satisfies the association property 
$(f\star g)\star h=f\star(g\star h)$, 
which limits forms of ${\cal F}^{n\ge 1}_A$ and 
$\beta_{n\ge 1}$. 
We note that the association property is 
necessary for the existence of the inverse 
with respect to the star product. 
For example, the inverse of the Lagrangian is a Green function, 
which always exists as $\psi\psi^\dagger$ with a wave function $\psi$. 

Now, we define a $p$-cochain space
${\cal C}^p:={\rm Hom}(V_f^{\otimes p},V_f)$ with 
$V^{\otimes p}_f\equiv\underbrace{V_f\otimes V_f\otimes\cdots\otimes
V_f}_p$, where 
$V_f$ represents a function space such as the Wigner space ${\cal W}$; 
we define a multi-vector space 
${\cal T}^k:=\Gamma({\cal M},\bigwedge^kTM)$, 
where ${\cal M}$ represents a manifold such as a classical phase space
(dimension $d$), $TM:=\bigcup_{p\in{\cal M}}T_pM$ 
denotes a tangent vector bundle with a tangent vector space 
$T_pM\equiv\{\sum_i^da^i(x)\partial_{x^i}\}$ at $p\in{\cal M}$ 
($x$ is a coordinate at $p$; $a_i$ represents a certain coefficient), 
$\bigwedge^k$ denotes a $k$-th completely antisymmetric tensor product, 
(for example, $\partial_i\wedge\partial_j=
\frac{1}{2!}(\partial_i\otimes\partial_j-
\partial_j\otimes\partial_i)\in \bigwedge^2 TM$),
and $\Gamma$ represents the section; 
for example, $\Gamma({\cal M},TM)$ 
is defined as a set of tangent vector 
at each position $p\in{\cal M}$. 
The Poisson bracket
$\{f,g\}\equiv\alpha(f\otimes g)
:=\alpha^{ij}(x)
(\partial_i\wedge\partial_j)(f\otimes g)$ 
is element of ${\cal T}^1$, 
where $\alpha^{ij}=-\alpha^{ji}$ is called the Poisson
structure $(i,j=1,2,\cdots,d)$. 

\begin{figure}[t]
\begin{center}
\includegraphics[width=12cm,clip]{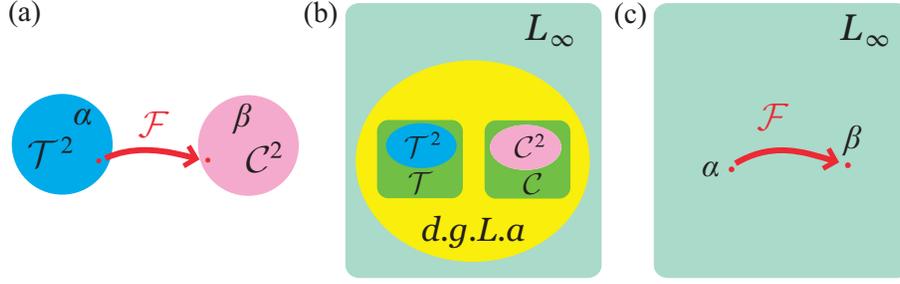}
\end{center}
\caption{
Steps of the derivation of the deformation quantization. 
(a): The image of the deformation quantization, which is the map from
 ${\cal T}^2$ with the Jacobi identity to ${\cal C}^2$ with the
 association property. 
(b): Enlargement of algebras. 
The two vector space ${\cal T}^2$ and two cochain space ${\cal C}^2$
 generalize to multi-vector space ${\cal T}$ and cochain space ${\cal C}$, respectively. 
These spaces are compiled in the d.g.L.a; finally, 
$L_\infty$ algebra is introduced by using the d.g.L.a.
(c): The deformation quantization is redefined as the map on the
 $L_\infty$ algebra.}\label{fig:DQ}
\end{figure}

The deformation quantization is the constitution method of higher order
cochains $\beta_{n\ge 2}\in{\cal C}^2$ from the Poisson bracket 
$\alpha\in{\cal T}^2$. 
In other words, the deformation quantization is the following map 
${\cal F}$: 
\begin{eqnarray}
 {\cal F}:~&&{\cal T}^2\rightarrow {\cal C}^2\nonumber\\
 &&\alpha\mapsto\beta\equiv\sum_{n\ge 1}\nu^n\beta_n,
\end{eqnarray}
where $\alpha$ satisfies the Jacobi identity
and $\beta$ satisfies the association property, 
as shown in Fig.~\ref{fig:DQ}(a). 

In the following sections, we will generalize the two-cochain ${\cal C}^2$
and the second order differential operator ${\cal T}^2$ to 
the so-called $L_\infty$ algebra 
(the definition is given in section~\ref{sub:L_infty}). 
In the section \ref{sec:CE}, 
we will introduce the two-cochain ${\cal C}^2$
and second order differential operator ${\cal T}^2$, 
and the $p$-cochain ${\cal C}^p$ and
$k$-th order differential operator ${\cal T}^k$. 
We will show that 
these operators satisfy certain conditions, and 
${\cal C}^p$ and ${\cal T}^k$ are embedded in a
differential graded Lie algebra (d.g.L.a) 
(the definition is shown in section~\ref{sec:CE}). 
Moreover, in section \ref{sub:L_infty}, 
the d.g.L.a will be embedded in 
the $L_\infty$ algebra 
(see Fig.~\ref{fig:DQ}(b)).
In the $L_\infty$ algebra, the Jacobi identity and the association
property are compiled in the following equation 
\begin{equation}
 Q(e^\gamma)=0,\label{eq:QE}
\end{equation}
where $\gamma=\alpha~{\rm or}~\beta$, and $Q$ is called the
codifferential operator,
which will be introduced in section \ref{sub:L_infty}.  
Namely, in the $L_\infty$ algebra, the deformation quantization is a map
from $\alpha\in{\cal T}^2$ to $\beta\in{\cal C}^2$ 
holding the solution of Eq.~(\ref{eq:QE})
(Figure~\ref{fig:DQ}(c)).
Such a map is uniquely determined in the $L_\infty$ algebra. 

In this paper, we will identify the tensor product $\otimes$ with the
direct product $\times$, i.e., 
$V_1\bigotimes V_2\sim V_1\times V_2$: 
$f\otimes g\sim (f,g)$ with $f\in V_1$ and $g\in V_2$ 
($a\sim b$ denotes that $a$ and $b$ are identified; 
$(f,g)$ represents the ordered pair, i.e., it is a set of 
$f$ and $g$, and $(a,b)\neq(b,a)$). 

\subsubsection{Cohomology equation}\label{sec:CE}
From Eq.~(\ref{eq:DQ0}), 
the association property is given by the following equation:
\begin{eqnarray}
\sum_{\substack{i+j=m \\ i,j\geq 0}}
\beta_i(\beta_j(f,g),h))=
\sum_{\substack{i+j=m \\ i,j\geq 0}}
\beta_i(f,\beta_j(g,h))\label{eq:AP}
\end{eqnarray}
with $\beta_0(f,g)\equiv f\cdot g$. 
(The symbol ``$\cdot$'' represents the usual commutative product, 
and $\beta_j\in{\cal C}^2,~j=0,1,\cdots$.) 
Because $\beta_1$ is the Poisson bracket, 
which is bi-linear differential operator, 
we define $\beta_j(\in{\cal C}^2,~j=2,3,\cdots)$ 
as a differential operator on a manifold
${\cal M}$; 
moreover, we also assume that $p$-cochains are 
differential operators and products of functions. 

Here, ${\cal A}$ and ${\cal C}^k({\cal A};{\cal A})$ represent 
a space of smooth functions on a manifold ${\cal M}$ and 
a space of multilinear differential maps from
${\cal A}^{\otimes k}$ to ${\cal A}$, respectively. 
Degree of $\beta^k\in{\cal C}^k({\cal A};{\cal A})$ is defined by
\begin{eqnarray}
 {\rm deg}(\beta^k):=
k\ \ \ {\rm for}~k\geq 2.
\end{eqnarray}
Now, we introduce a coboundary operator 
$\partial_C:{\cal C}^k({\cal A};{\cal A})
\rightarrow {\cal C}^{k+1}({\cal A};{\cal A})$
~\cite{1945,springerlink:10.1007/BFb0084073};
\begin{eqnarray}
 (\partial_C \beta^k)(\underbrace{f_0,\cdots,f_k}_{k+1})&:=&
f_0\beta^k(\underbrace{f_1,\cdots,f_k}_{k})+
\sum_{r=1}^k(-1)^r
\beta^k(\underbrace{f_0,\cdots,f_{r-1}\cdot f_r,\cdots f_k}_{k})\nonumber\\
&&\ \ \ \ \ \ +(-1)^{k-1}
\beta^k(\underbrace{f_0,\cdots,f_{k-1}}_{k})f_k
\end{eqnarray}
with 
$\beta^k\in{\cal C}^k({\cal A};{\cal A})$; 
note that $\partial_C^2=0$, and thus, 
$\partial_C$ is the boundary operator. 
The Gerstenhaber bracket is defined as 
$[~,~]_C:{\cal C}^k({\cal A};{\cal A})\otimes {\cal C}^{k^\prime}
({\cal A};{\cal A})\rightarrow 
{\cal C}^{k+k^\prime-1}({\cal A};{\cal A})$~\cite{Grestenhaber}: 
\begin{eqnarray}
&&[\beta^k,\beta^{k^\prime}]_C
(\underbrace{f_0,f_1,\cdots,f_{k+k^\prime-2} }_{k+k^\prime-1})
\nonumber\\
&&:=\sum^{k-1}_{r=0}(-1)^{r(k^\prime-1)}
\beta^k(\underbrace{f_0,\cdots ,f_{r-1},
\beta^{k^\prime}(f_r,\cdots,f_{r+k^\prime-1}),f_{r+k^\prime},\cdots,f_{k+k^\prime-2}}_{k})\nonumber\\
&&-\sum_{r=0}^{k^\prime-1}(-1)^{(k-1)(r+k^\prime-1)}
\beta^{k^\prime}(\underbrace{f_0,\cdots,f_{r-1},
\beta^k(f_r,\cdots,f_{r+k-1}),f_{r+k},\cdots,f_{k+k^\prime-2}}_{k^\prime}),\nonumber\\
\end{eqnarray}
where $\beta^k\in {\cal C}^k({\cal A};{\cal A})$ and 
$\beta^{k^\prime}\in{\cal C}^{k^\prime}({\cal A};{\cal A})$. 
Note that $\partial_C^2=0$, and thus, 
$\partial_C$ is the boundary operator. 

By using the coboundary operator and the Gerstenhaber bracket,
Eq.~(\ref{eq:AP}) is rewritten as 
\begin{equation}
 \partial_C \beta_m=-\frac{1}{2}\sum_{\substack{i+j=m \\ i,j\geq 0}}
[\beta_i,\beta_j]_C\label{eq:CE}
\end{equation}
with $\beta_j\in{\cal C}^2({\cal A};{\cal A})$ ($j=1,2,\cdots$). 
For example, Eq.~(\ref{eq:AP}) for $m=0,1,2$ is given as:
\begin{eqnarray}
\begin{array}{cc}
(f\cdot g)\cdot h = f\cdot (h\cdot g) & {\rm for}~m=0,\label{eq:CE0}\\
\{f\cdot g,h\} + \{f,g\}\cdot h
=\{f,g\cdot h\} + f\cdot \{g,h\} & {\rm for}~m=1,\\
\beta_2(f\cdot g,h) + \{\{f,g\},h\} + \beta_2(f,g)\cdot h
=\beta_2(f,g\cdot h) + \{f,\{g,h\}\} + f\cdot \beta_2(g,h) 
& {\rm for}~m=2,
\end{array}
\end{eqnarray}
where we have used $\beta_0(f,g):=f\cdot g$ and
$\beta_1(f,g)\equiv\{f,g\}$. 
The coboundary operator for $\beta\in{\cal C}^2({\cal A};{\cal A})$
is given by:
\begin{equation}
 (\partial_C \beta)(f,g,h)=
f\cdot \beta(g,h)-\beta(f\cdot g,h)
+\beta(f,g\cdot h)-\beta(f,g)\cdot h;
\end{equation}
moreover, the Gerstenhaber bracket in terms of 
$\beta_i,\beta_j\in{\cal C}^2({\cal A};{\cal A})$ is given by 
\begin{equation}
 [\beta_i,\beta_j]_C(f,g,h)=
\beta_i(\beta_j(f,g),h)-\beta_i(f,\beta_j(g,h))+
\beta_j(\beta_i(f,g),h)-\beta_j(f,\beta_i(g,h)).\label{eq:GB}
\end{equation}
Using the above Eqs.~(\ref{eq:CE0}-\ref{eq:GB}), 
we can check the equivalence between Eq.~(\ref{eq:AP}) and
Eq.~(\ref{eq:CE}). 

Equation (\ref{eq:CE}) is called the cohomology equation, and the star
product is constructed by using solutions of the cohomology equation. 
If we add Eq.~(\ref{eq:CE}) with respect to $m=0,1,2,\cdots$, 
we obtain the following equation:
\begin{equation}
 \partial_C \beta+\frac{1}{2}[\beta,\beta]_C=0\label{eq:MCC}
\end{equation}
with $\beta\equiv\sum_{j=0}^\infty \beta_j$; 
$\beta,\beta_j\in{\cal C}^2({\cal A};{\cal A})$, $j=0,1,2,\dots$.

Here, we identify the vector fields $\partial_i,\partial_j\in TM$ with
anti-commuting numbers $\tilde{\eta}_i,\tilde{\eta}_j$ 
($\tilde{\eta}_i\tilde{\eta}_j=-\tilde{\eta}_j\tilde{\eta}_i$), 
$i,j=1,2,\dots,d$; 
thus the Poisson bracket $\alpha^{ij}(\partial_i\wedge\partial_j)/2$ 
is rewritten by 
$\alpha=\alpha^{ij}\tilde{\eta}_i\tilde{\eta}_j/2$. 
Now, we define the Batalin-Vilkovisky (BV) bracket:
\begin{equation}
 [\alpha_1,\alpha_2]_{\rm BV}:=-\sum_{i=1}^d\left(
\frac{\alpha_1\overleftarrow{\partial}}{\partial x^i}
\frac{\overrightarrow{\partial}\alpha_2}{\partial \tilde{\eta}_i}-
\frac{\alpha_1\overleftarrow{\partial}}{\partial \tilde{\eta}_i}
\frac{\overrightarrow{\partial}\alpha_2}{\partial x^i}
\right)
\end{equation}
with $\alpha_1,\alpha_2\in{\cal T}^2$.
By using the BV bracket, the Jacobi identity is rewritten as
\begin{equation}
 \partial_{\rm BV} \alpha +\frac{1}{2}[\alpha,\alpha]_{\rm BV}=0,\label{eq:BVE}
\end{equation}
with $\alpha_1,\alpha_2\in{\cal T}^2$;
for $\alpha=\alpha^{ij}\tilde{\eta}_i\tilde{\eta}_j$, 
$\alpha\overleftarrow{\partial}/\partial x^l=
\overrightarrow{\partial}\alpha
/\partial x^l:=(\partial_{x^l}\alpha^{ij})
\tilde{\eta}_i\tilde{\eta}_j$
and
$\overrightarrow{\partial}\alpha/\partial\tilde{\eta}_l
=-\alpha\overleftarrow{\partial}/\partial\tilde{\eta}_l
:=\alpha^{ij}(\delta_{il}\tilde{\eta}_j-\tilde{\eta}_i\delta_{jl})$. 
By using the BV bracket, the Jacobi identity is rewritten as
\begin{equation}
 \partial_{\rm BV} \alpha +\frac{1}{2}[\alpha,\alpha]_{\rm BV}=0,\label{eq:BVE}
\end{equation}
where $\partial_{\rm BV}\equiv 0$, i.e., $\partial_{{\rm BV}}^2=0$.

Now, we generalize the differential $\partial_{\rm BV}$ and 
BV-bracket $[~,~]_{\rm BV}$ for 
$\alpha^k\in {\cal T}^k$ and $\alpha^{k^\prime}\in{\cal T}^{k^\prime}$ as follows
(${\cal T}^k\equiv \Gamma({\cal M},\bigwedge^k TM)$):
\begin{eqnarray}
&&\partial_{\rm BV}:{\cal T}^k\rightarrow{\cal T}^{k+1}\nonumber\\
&&\ \ \ \ \ \ \ \ \partial_{\rm BV}:=0,\\
&&[~,~]_{\rm BV}:{\cal T}^k\bigotimes
{\cal T}^{k^\prime}\rightarrow {\cal T}^{k+k^\prime-1}\nonumber\\
&&[\alpha^k,\alpha^{k^\prime}]_{\rm BV}:=-
\sum_{i=1}^d\left(\frac{\alpha^k\overleftarrow{\partial}}{\partial\tilde{\eta_i}}
\frac{\overrightarrow{\partial}\alpha^{k^\prime}}{\partial x^i}-
\frac{\alpha^k\overleftarrow{\partial}}{\partial x^i}
\frac{\overrightarrow{\partial}\alpha^{k^\prime}}
{\partial \tilde{\eta}_i}\right)
\end{eqnarray}
with 
$\alpha^k=
(\alpha^k)^{i_1,\cdots,i_k}(x)\eta_{i_1}\wedge\cdots\wedge\eta_{i_k}
\sim(\alpha^k)^{i_1,\cdots,i_k}(x)\tilde{\eta}_{i_1}\cdots\tilde{\eta}_{i_k}$, 
and 
$\alpha^{k^\prime}=
(\alpha^{k^\prime})^{i_0,\cdots,i_{k^\prime}}(x)
\eta_{i_0}\wedge\cdots\wedge\eta_{i_{k^\prime}}
\sim(\alpha^{k^\prime})^{i_0,\cdots,i_{k^\prime}}(x)
\tilde{\eta}_{i_0}\cdots\tilde{\eta}_{i_{k^\prime}}$; 
degree of $\alpha\in{\cal T}^k$ is defined by
\begin{equation}
 {\rm deg}(\alpha)=k-1,~\alpha\in{\cal T}^k.
\end{equation}

The cochain algebra is defined by 
the set of the differential operator $\partial_C$, 
the Gerstenhaber bracket $[~,~]_C$ and 
${\cal C}:=\bigoplus_{k=2}^\infty{\cal C}^k$, i.e.,
$(\partial_C,[~,~]_C,{\cal C})$; 
in addition, the multi-vector algebra is defined by 
the set of the differential operator $\partial_{\rm BV}:=0$, 
BV bracket $[~,~]_{\rm BV}$ and 
${\cal T}:=\bigoplus_{k=1}^\infty{\cal T}^k$, i.e.,
$(\partial_{\rm BV},[~,~]_{\rm BV},{\cal T})$. 
The cochain algebra and the multi-vector algebra satisfy 
the following common relations:
\begin{equation}
 \partial^2=0,\label{eq:dgla1}
\end{equation}
\begin{equation}
 \partial[\gamma_1,\gamma_2]=
[\partial\gamma_1,\gamma_2]+
(-1)^{{\rm deg}(\gamma_1)}
[\gamma_1,\partial\gamma_2],\label{eq:dgla2}
\end{equation}
\begin{equation}
 [\gamma_1,\gamma_2]=
-(-1)^{{\rm deg}(\gamma_1){\rm deg}(\gamma_2)}
[\gamma_2,\gamma_1]\label{eq:dgla3}
\end{equation}
\begin{equation}
 [\gamma_1,[\gamma_2,\gamma_3]]
+(-1)^{{\rm deg}(\gamma_3)({\rm deg}(\gamma_1)+
{\rm deg}(\gamma_2))}[\gamma_3,[\gamma_1,\gamma_2]]+
(-1)^{{\rm deg}(\gamma_1)({\rm deg}(\gamma_2)+
{\rm deg}(\gamma_3))}[\gamma_2,[\gamma_3,\gamma_1]]=0\label{eq:dgla4}
\end{equation}
with $\gamma_1,\gamma_2,\gamma_3\in{\cal G}
\equiv({\cal C}\ {\rm or}\ {\cal T})$,
$\partial\equiv\partial_{(C\ {\rm or}\ BV)}$, and 
$[~,~]\equiv[~,~]_{(C\ {\rm or}\ BV)}$. 
Therefore, the two algebra can be compiled in the so-called the
differential graded Lie algebra (d.g.L.a) $(\partial,[~,~],{\cal G})$, 
where ${\cal G}:=\bigoplus_{k=1}^\infty{\cal G}^k$ is a graded $k$-vector
space with ${\cal G}^k$ has a degree ${\rm deg}(x)\in\mathbb{Z}$
($x\in{\cal G}^k$; $\mathbb{Z}$ is the set of integers), 
and d.g.L.a. has the linear operator $\partial$ and the bi-linear operator $[~,~]$:
\begin{eqnarray}
 &&\partial:~{\cal G}^k\rightarrow {\cal G}^{l},~
x_k\in{\cal G}^k,~x_l\in{\cal G}^l,\nonumber\\
 &&\ \ \ \ \ \ 
{\rm deg}(\partial x_k)={\rm deg}(x_k)+1={\rm deg}(x_l),\\
 &&[~,~]:~{\cal G}^k\bigotimes{\cal G}^l
 \rightarrow{\cal G}^m,~
x_k\in{\cal G}^k,~x_l\in{\cal G}^l,~x_m\in{\cal G}^m,
\nonumber\\
 &&\ \ \ \ \ \ 
{\rm deg}([x_k,x_l])={\rm deg}(x_k)+{\rm deg}(x_l)={\rm deg}(x_m),
\end{eqnarray}
where $\partial$ and $[~,~]$ satisfy 
Eqs.~(\ref{eq:dgla1}),~(\ref{eq:dgla2}),~(\ref{eq:dgla3}) and (\ref{eq:dgla4}).
In d.g.L.a., Eqs.~(\ref{eq:MCC}) and (\ref{eq:BVE}) are
compiled in the so-called Maurer-Cartan equation~\cite{Cartan}:
\begin{equation}
 \partial\gamma+\frac{1}{2}[\gamma,\gamma]=0\label{eq:MC}
\end{equation}
with $\gamma\in{\cal G}$. 
Therefore, the deformation quantization ${\cal F}$ is a map:
\begin{eqnarray}
 &&{\cal F}:{\cal G}\rightarrow {\cal G},~
\gamma_1\mapsto \gamma_2,\nonumber\\
 &&\partial\gamma_i+\frac{1}{2}[\gamma_i,\gamma_i]=0,~i=1,2.
\end{eqnarray}
Namely, the deformation quantization is a map holding a solution of the
Maurer-Cartan equation~(\ref{eq:MC}). 
In the section~\ref{sub:L_infty}, we will introduce a $L_\infty$ algebra,
and will redefine the deformation quantization; in the $L_\infty$
algebra, the Maurer-Cartan equation~(\ref{eq:MC}) is rewritten as 
$Q(e^\gamma)=0$ ($Q$ and $e^\gamma$ will be defined in \ref{sub:L_infty}).

\subsubsection{$L_\infty$ algebra}\label{sub:L_infty}
Now we define a commutative graded coalgebra ${\cal C}(V)$. 

First, we define a set $({\cal V},\triangle,\tau,Q)$, where 
${\cal V:=}\bigoplus_{n=1,2,\cdots}V^{\otimes n}$ 
with a graded $k$-vector space $V^{\otimes n}$ ($n=1,2,\dots$), 
$\triangle$ and $Q$ represent the coproduct 
and codifferential operator, respectively;
moreover, $\tau$ denotes cocommutation (definition is given later).
The coproduct, cocommutation and codifferential operator satisfy the following equations: 
\begin{eqnarray}
&&(\triangle\otimes {\rm id})\circ\triangle=({\rm id}\otimes\triangle)\circ\triangle,\\
&&\tau\triangle=\triangle,\\
&&\triangle\circ Q=({\rm id}\otimes Q+Q\otimes {\rm id})\circ\triangle,\\
&&\tau(x\otimes y):=(-1)^{{\rm deg}_{\rm co}(x){\rm deg}_{\rm co}
(y)}y\otimes x,
\end{eqnarray}
with ${\rm deg}_{\rm co}(x):={\rm deg}(x)-1$,
where $x\in V^{\otimes{\rm deg}(x)}$ and 
$y\in V^{\otimes{\rm deg}(y)}$. 
$Q$ represents a codifferential operator adding one degree: 
$Q\in{\rm Hom}(V^{\otimes m},V^{\otimes(m+1)})$ with 
${\rm deg}_{\rm co}(Q(x))={\rm deg}_{\rm co}(x)+1$ for
$x\in V^{\otimes m},~{}^\exists m\in\mathbb{Z}_+$ 
(the explicit form of $Q$ is given later; 
$\mathbb{Z}_+:=\{i~|~i>0,~i\in\mathbb{Z}\}$). 

By using $\tau$, we define the commutative graded coalgebra 
${\cal C}(V)$ from $({\cal V},\triangle,\tau,Q)$; 
the identify relation $\sim$ is defined as 
$x\otimes y\sim(-1)^{{\rm deg}_{\rm co}(x){\rm deg}_{\rm co}(y)}y\otimes x$,
i.e., $x\otimes y$ and 
$(-1)^{{\rm deg}_{\rm co}(x){\rm deg}_{\rm co}(y)}y\otimes x$ are identified.  
Now, we define the commutative graded tensor algebra:
\begin{equation}
{\cal C}(V):={\cal V}/\sim~
\equiv\{[x]|x\in{\cal V}\},
\end{equation} 
where $[x]=\left\{y~|~y\in{\cal V},~x\sim y\right\}$, 
and ${\rm deg}_{\rm co}(x_1\otimes x_2\otimes\cdots\otimes x_n)=
{\rm deg}_{\rm co}(x_1)+{\rm deg}_{\rm co}(x_2)
+\cdots+{\rm deg}_{\rm co}(x_n)$ with 
$x_1\otimes x_2\otimes\cdots \otimes x_n\in 
V^{\otimes{\rm deg}_{\rm co}(x_1)}\otimes
V^{\otimes{\rm deg}_{\rm co}(x_2)}
\otimes\cdots\otimes V^{\otimes{\rm deg}_{\rm co}(x_n)}$; 
a product in ${\cal C}(V)$ is defined by $xy:=[x\otimes y]$. 
Namely, in ${\cal C}(V)$, 
\begin{equation}
 x_1x_2\cdots x_ix_{i+1}\cdots x_n=
(-1)^{{\rm deg}_{\rm co}(x_i){\rm deg}_{\rm co}(x_{i+1})}x_1x_2\cdots 
x_{i+1}x_i\cdots x_n
\end{equation}
with $n\geq 2$. 
(Let us recall that the derivation of the exterior algebra from the tensor
space; ${\cal V}$ and ${\cal C}(V)$ correspond to the tensor space
and the exterior algebra, respectively.)

Moreover, in the case that $Q^2=0$, the commutative graded coalgebra
${\cal C}(V)$ is called the $L_\infty$ algebra. 
For the $L_\infty$ algebra, the coproduct and codifferential operator are
uniquely determined by using multilinear operators:
\begin{eqnarray}
 &&l_k: (V^{\otimes k}\in{\cal C}(V)))\rightarrow V\in{\cal C}(V)\\
 &&{\rm deg}_{\rm co}(l_k(x_1\cdots x_k))=
{\rm deg}_{\rm co}(x_1)+\cdots+{\rm deg}_{\rm co}(x_k)+1
\end{eqnarray}
as follows: 
\begin{eqnarray}
\triangle(x_1\cdots x_n)&=&
\sum_\sigma\sum_{k=1}^{n-1}
\frac{\varepsilon(\sigma)}{k!(n-k)!}
(x_{\sigma(1)}\cdots x_{\sigma(k)})\otimes
(x_{\sigma(k+1)}\cdots x_{\sigma(n)}),\\ 
Q&=&\sum_{k=1}^\infty Q_k,\\
Q_k(x_1\cdots x_n)&=&
\sum_\sigma\frac{\varepsilon(\sigma)}{k!(n-k)!}
l_k(x_{\sigma(1)}\cdots x_{\sigma(k)})\otimes 
x_{\sigma(k+1)}\otimes\cdots\otimes x_{\sigma(n)},\nonumber\\
\end{eqnarray}
where $\varepsilon(\sigma)$ represents a sign with a replacement
$\sigma:x_1x_2\cdots x_n
\mapsto x_{\sigma(1)}x_{\sigma(2)}\cdots x_{\sigma(n)}$. 
From the condition $Q^2=0$, we can identify $(l_1,~l_2)$ with
$(\partial,~[\ ,\ ])$ in d.g.L.a. 
If we put $l_3=l_4=\cdots=0$, $Q(e^\alpha)=0$ for $\alpha\in V$ is
equal to the Maurer-Cartan equation Eq.~(\ref{eq:MC}) in d.g.L.a~\cite{Kontsevich}, 
where
\begin{equation}
 e^\alpha\equiv 1+\alpha+\frac{1}{2!}\alpha\otimes\alpha+\cdots
\end{equation}
with $\alpha^{\otimes n}\otimes 1\equiv 1\otimes\alpha^{\otimes
n}\equiv \alpha^{\otimes n}~{\rm
for}~n=1,2,\cdots$. 
Therefore, the deformation quantization is a map:
\begin{eqnarray}
 {\cal F}&:&~{\cal C}(V)\rightarrow 
{\cal C}(V),\nonumber\\
 &&\ \ \gamma_1\mapsto\gamma_2
\end{eqnarray} 
with
\begin{equation}
 Q(e^{\gamma_i})=0,~i=1,2.
\end{equation}

To constitute such a map ${\cal F}$, 
we introduce the $L_{\infty}$ map ${\mathscr F}$,  
which is defined as the following map holding degrees of coalgebra:
\begin{eqnarray}
 {\mathscr F}:&&~{\cal C}(V)\rightarrow{\cal C}(V),~
v_1,v_2\in{\cal C}(V),\nonumber\\
 &&v_1\mapsto v_2,\nonumber\\
 &&{\rm deg}_{\rm co}(v_1)={\rm deg}_{\rm co}(v_2);
\end{eqnarray}
moreover, the $L_\infty$ map satisfies the following equations:
\begin{equation}
 \triangle\circ{\mathscr F}=({\mathscr F}\otimes
{\mathscr F})\circ\triangle,
\end{equation}
\begin{equation}
 Q\circ{\mathscr F}={\mathscr F}\circ Q.
\end{equation}

A form of such a map is limited as~\cite{Kontsevich}:
\begin{eqnarray}
 &&{\mathscr F}={\mathscr F}^1+\frac{1}{2!}{\mathscr F}^2+
\frac{1}{3!}{\mathscr F}^3+\cdots,\\
 &&{\mathscr F}^l:{\cal C}(V)\rightarrow 
V^{\otimes l}(\subset{\cal C}(V))\nonumber\\
 &&{\mathscr F}^l(x_1\cdots x_n)=
\sum_\sigma\sum_{\substack{n_1,\cdots,n_l\geq 1 \\
 n_1+\cdots+n_l=n}}\frac{\varepsilon(\sigma)}
{n_1!\cdots n_l!}\nonumber\\
&&\ \ \ \ \ \ \ \ \ \ \ \ \ \ \ \ \ \ \ \ \ \ 
\cdot{\mathscr F}_{n_1}(x_{\sigma(1)}\cdots x_{\sigma(n_1)})
\otimes\cdots\otimes{\mathscr F}_{n_l}
(x_{\sigma(n-n_l+1)}\cdots x_{\sigma(n)}),
\end{eqnarray}
where ${\mathscr F}_n$ is a map from ${\cal C}(V)$ to
$V(\subset{\cal C}(V))$ holding degrees;
\begin{eqnarray}
 {\mathscr F}_n:\ \ \ \ \ \ \ \ \ \ \ \ \ \ \ \ \ 
V^{\otimes n}(\subset {\cal C}(V))&\rightarrow &
V(\subset {\cal C}(V))\nonumber\\
x_1\otimes\cdots\otimes x_n&\mapsto &x^\prime,\nonumber\\
{\rm deg}_{\rm co}(x_1)+\cdots+{\rm deg}_{\rm co}(x_n)&=&
{\rm deg}_{\rm co}(x^\prime). 
\end{eqnarray}

Here we define 
$\beta:=\sum_{n=1}^\infty \frac{1}{n!}
{\mathscr F}_n(\alpha\cdots\alpha)$, 
which satisfies 
${\mathscr F}(e^\alpha)=e^\beta$. 
The map ${\mathscr F}$ holds solutions of Maurer-Cartan equations 
$Q\left(e^\alpha\right)=0$
and $Q\left(e^\beta\right)=0$; 
from
\begin{equation}
 Q(e^\beta)\equiv Q\circ{\mathscr F}\left(e^\alpha\right)
\end{equation}
and the definition of the $L_\infty$ map: 
$Q\circ{\mathscr F}={\mathscr F}\circ Q$, 
we obtain the following equation:
\begin{equation}
 Q(e^\beta)={\mathscr F}\circ Q\left(e^\alpha\right)=0,
\end{equation}
which means that the $L_\infty$ map transfers 
a solution of the Maurer-Cartan equation from another solution. 

Now, we return to the deformation quantization. 
The multi-vector space ${\cal T}$, 
is embedded in $C(V)$; 
$C(V)=({\cal T},\triangle_{\cal T},\tau,Q_{\cal T})$, where 
$\triangle_{\cal T}(x_1x_2):=x_1\wedge x_2$ for $x_1,x_2\in{\cal T}$, 
$(Q_{\cal T})_1:=\partial_{\rm BV}\equiv 0$, 
$(Q_{\cal T})_2:=[~,~]_{\rm BV}$, 
and $(Q_{\cal T})_l:=0$ for $l=3,4,\dots$; 
$\tau$ replaces the wedge product ``$\wedge$'' with the product
``$\cdot$''. 
For the cochain space ${\cal C}$, 
it is also embedded in $C(V)$; 
$C(V)=({\cal C},\triangle_{\cal C},\tau,Q_{\cal C})$, where 
$\triangle_{\cal C}(x_1x_2):=x_1\wedge x_2$ for $x_1,x_2\in{\cal C}$,
$(Q_{\cal C})_1:=\partial_C$, $(Q_{\cal C})_2:=[~,~]_C$, and 
$(Q_{\cal C})_l:=0$ for $l=3,4,\dots$.

The star product is given by
$f\star g=f\cdot g+\beta(f\otimes g)$, which is identified as the map 
${\mathscr F}_0+{\mathscr F}^1$ with ${\mathscr F}_0:=\mu\circ$. 

Here we summarize the main results of the succeeding sections 
without explaining their derivations. 
The map ${\mathscr F}^1$ is given by a path integral of 
a topological field theory 
having super fields: 
$X:=(X^1,\dots,X^{N})$; 
and scalar fields: $\psi:=(\psi^1,\dots,\psi^{N})$, 
$\lambda:=(\lambda^1,\dots,\lambda^N)$,
and $\gamma:=(\gamma^1,\dots,\gamma^{N})$; 
and one-form fields: 
$\theta:=(\theta^1,\theta^2,\dots,\theta^m)$, 
$A:=(A_1,\dots,A_{N})$, 
$A^+:=(A^{+1},\dots,A^{+N})$, 
and $\eta:=(\eta^{1},\dots,\eta^{N})$; 
and Grassmann fields $c_i:=(c^1,\dots,c^{N})$;
on a disk 
$\Sigma=\{z~|~z=u+{\rm i}v,~u,v\in\mathbb{R},~v\geq 0\}$
~\cite{Kontsevich,Cattaneo,Izawa}. 
These fields are defined in
section~\ref{sec:TopologicalStringTheory}. 
Using these fields, the map ${\mathscr F}_n: 
V^{\otimes n}\rightarrow V$ 
is given as follows:
\begin{eqnarray}
 {\mathscr F}_n(\alpha_1,\cdots,\alpha_n)
(f_1\otimes \cdots\otimes f_m)(x)=
\int e^{\frac{{\rm i}}{\hbar}S^0_{\rm gh}}
\frac{{\rm i}}{\hbar}S_{\alpha_1}\cdots
\frac{{\rm i}}{\hbar}S_{\alpha_n}{\cal O}_x
(f_1,\dots,f_m)
\end{eqnarray}
for any function $f_1,\dots,f_m$, which depend on $x$; 
in this paper, 
$x$ represents the coordinate in the classical phase space.
Here $\alpha_1,\alpha_2,\dots,\alpha_n\in V$, and 
$m$ is defined by ${\rm deg}_{\rm co}(\alpha_i)+2$, 
which is common and independent of $i$ ($i=1,2,\dots,n$). 

The operator ${\cal O}_x$ is defined as
\begin{eqnarray}
 {\cal O}_x(f_1,\dots,f_m)&:=&
\int \left[
f_1(X(t_1,\theta_1))\cdots f_m(X(t_m,\theta_m))\right]
\delta_x(\psi(\infty))\\
&\equiv&
\int_{1=t_1>t_2>\cdots>t_m=0}
f_1(\psi(t_1))\prod_{k=2}^{m-1}
\partial_{i_k}\left[
f(\psi(t_{k}))A^{+i_k}(t_k)
\right]
f_m(\psi(0))\delta_x(\psi(\infty))\label{eq:PathO}
\end{eqnarray}
for $m$, 
$\delta_x(\psi(t)):=\prod_{i=1}^d
\delta(\psi^i-x^i)\gamma^i(t)$, and 
$t\in\partial\Sigma$, 
where 
\begin{equation}
 S^0_{\rm gh}:=
\int_\Sigma
\left[
A_i\wedge d\psi^i-*_Hd\gamma^i\wedge dc_i
-\lambda^id*_HA_i
\right]
\end{equation}
with a Hodge operator $*_{\rm H}:\wedge^{k}\rightarrow\wedge^{2-k}$, 
($k=0,1,2$); 
we will introduce the explicit definition in section~\ref{sec:PerturbationTST}.
Moreover, for 
$\alpha_r:=\alpha^{i_1,\cdots,i_{n_r}}_r(X)\partial_{i_1}\wedge
\cdots\wedge\partial_{i_{n_r}}$ 
($n_r>1$ is an integer number; ${\rm deg}_{\rm co}=n_r-2$), 
\begin{eqnarray}
 S_{\alpha_r}:=\left.\left(
\int_\Sigma\int d^2\theta\frac{1}{n_r}\alpha_r^{i_1\cdots i_{n_r}}(X)
\eta_{i_1}\cdots \eta_{i_{n_r}}\right)
\right|_{\Phi^*=\partial\varphi},
\end{eqnarray}
where the subscript $\Phi^*=\partial\varphi$ means 
that the fields ($X,\eta,A^+$) go to ($\psi,A,0$).
These results lead to the diagram technique in
section~\ref{sec:DiagramRules} and the explicit expression of the star
product in section~\ref{sec:TwistedElement}. 

\subsection{Topological string
  theory}\label{sec:TopologicalStringTheory}
In this section, we expound the fields: 
$A$, $\psi$, $c$, $\gamma$, $\lambda$, $\theta$, $\eta$, $A^{+}$, and $X$. 
The simplest topological string theory is defined the following 
action:
\begin{equation}
 S_0:=\int_{\Sigma}d^2\sigma\epsilon^{\mu\nu}A_{\mu,i}\partial_\nu\psi^i
=-\frac{1}{2}\int_\Sigma d^2\sigma\epsilon^{\mu\nu}F_{\mu\nu,i}\psi^i
\label{eq:TS}
\end{equation}
with local coordinates $\sigma=(\sigma^1,\sigma^2)$ on a disk $\Sigma$
(we consider that the disk is the upper-half plain in the complex one,
i.e., $\Sigma:=\{z~|~z=u+{\rm i}w;~w\ge 0;~u,w\in\mathbb{R}\}$),
where $A_{\mu,i}(\sigma)$ and $\psi^i(\sigma)$ are 
U(1) gauge fields and scalar fields, respectively; 
$F_{\mu\nu,i}(\sigma)$ is a gauge strength 
($\mu=1,2$ and $i=1,\dots,N$). 
The other fields $c$, $\gamma$, $\lambda$, $\theta$, $\eta$, and $A^+$ are 
introduced in section \ref{sec:GF}; 
we discuss the gauge fixing method using the so-called
BV-BRST formalism~\cite{Batalin198127,PhysRevD.28.2567}
(where the BV refers to Batalin and Vilkovisky; 
BEST refers to Becchi, Rouet, Stora and Tyutin). 
In section \ref{sec:CGC}, we discuss the gauge invariance of the path
integral, and introduce the SD operator. 
In section \ref{sec:PathIntegralTST}, we see that correspondence of
the deformation quantization and topological string theory. 

\subsubsection{Ghost fields and anti-fields}\label{sec:GF}

Here, we quantize the action (\ref{eq:TS}) using the path integral. 
Roughly speaking, the path integral is the Gauss integral around a
solution of an equation of motion. 
In many cases, a general action $S$ has no inverse. 
Therefore, we will add some extra fields, 
and obtain the action $S_{\rm gh}$ having inverse, 
which is called as the quantized action.

Now, we discuss a general field theory. 
We assume that a general action $S$ 
is a function of fields $\phi^i$, i.e.,
$S=S[\phi^i]$; 
each field $\phi^i$ is labeled by a certain integer number, 
which is called as a ghost number 
${\rm gh}(\phi^i)$ (it is defined below). 
$\phi_C^i$ denotes that the fields fixed on 
the solution of the classical kinetic equation: 
$\delta S_0/\delta \phi^i=0$, 
and the subscript of the fields represents a number of fields. 
Because the Gauss integral is an inverse of a Hessian,
a rank of the Hessian should be equal to the number of the fields. 
Here a Hessian is defined by:
\begin{equation}
 K[\phi^i,\phi^j]:=
 \frac{\overrightarrow{\delta}}{\delta\phi^i}S
 \frac{\overleftarrow{\delta}}{\delta\phi^j},
\end{equation}
where 
$\frac{\overrightarrow{\delta}}{\delta\phi^i}\phi^{j_1}\phi^{j_2}
\cdots\phi^{i_n}:=\delta^{j_1}_i\phi^{j_2}\cdots\phi^{j_2}
+(-1)^{\overline{j}_1\overline{i}}\phi^{j_1}\delta^{j_2}_i\cdots\phi^{j_n}_i
+\cdots+(-1)^{\overline{i}(\overline{j}_1+\overline{j}_2+\cdots+
\overline{j}_{(n-1))}}\phi^{j_1}\phi^{j_2}\cdots\phi^{j_{(n-1)}}\delta^{j_n}_i$, 
and 
$\phi^{i_1}\phi_{i_2}\cdots\phi^{i_n}
\frac{\overleftarrow{\delta}}{\delta\phi_j}:=
\phi^{i_1}\phi^{i_2}\cdots\phi^{i_{(n-1)}}\delta^{i_n}_j
+(-1)^{\overline{i}_n\overline{j}}
\phi^{i_1}\phi^{i_2}\cdots\delta^{i_{(n-1)}}_j\phi^{i_n}
+(-1)^{(\overline{i}_2+\cdots+\overline{i}_n)\overline{j}}
\delta^{i_1}_j\phi^{i_2}\cdots\phi^{i_n}$; 
for a boson $\phi^i$, $\overline{i}$ is a ghost number ${\rm gh}(\phi^i)$;
for a fermion $\phi^i$, $\overline{i}$ is ${\rm gh}(\phi^i)+1$. 

We define the rank of the Hessian $K$ and the number of the fields
$\phi^i$ by $\sharp K$ and $\sharp \phi^i$, respectively. 
Generally speaking, $\sharp K<\sharp \phi^i$, because an action has
some symmetries $\delta_R\phi^i:=R^i_j\phi^j$ 
with nontrivial symmetry generators $R^i_j$, 
where is satisfies the following equation:
\begin{equation}
 \frac{S\overleftarrow{\delta}}{\delta\phi^i}R^i_j=0
\end{equation}
with $R^i_j|_{\phi^k=\phi^k_c}\neq 0$. 
The nontrivial symmetry generator decrease 
the rank of Hessian from the number of fields. 
To define the path integral, 
we should add $(\sharp\phi^i-\sharp K)$ virtual fields
\cite{Batalin198127,PhysRevD.28.2567,Alfaro1993751}. 
The additional fields are called as ghost fields $\Phi^{\alpha_1}$ 
and antifields $\Phi^{*}_{\alpha_l}$ ($l=0,1$), 
and these fields are labeled by ghost numbers. 
For $\Phi^{\alpha_1}$, the ghost number is defined by 
${\rm gh}(\Phi^{\alpha_1}):=1$. 
The fields and ghost fields have antifields $\Phi_{\alpha_l}^*$. 
The antifields corresponding to $\phi\equiv \Phi^{\alpha_0}$ 
and $\Phi^{\alpha_1}$ are 
described as $\Phi^*_{\alpha_0}$ and $\Phi^*_{\alpha_1}$, respectively. 
The ghost number of $\Phi^*_{\alpha_l}$ is defined by 
${\rm gh}(\Phi^*_{\alpha_l})=-l-1$. 
Statistics of the anti-fields is opposite of fields, i.e., 
if the fields are fermions(bosons), the anti-fields are bosons(fermions).
(Here we only consider the so-called irreducible theory.
For a general theory, see references
\cite{Batalin198127,PhysRevD.28.2567,Alfaro1993751}.)

Using these fields, we will transform the action 
$S[\Phi^{\alpha_0}]\mapsto S_{\rm gh}[\Psi]$, 
where $\Psi:=(\Phi^{\alpha_l},\Phi^*_{\alpha_l})$ 
with $l=0,1$ and 
$\alpha_l\in\mathbb{Z}_+:=\{i~|~i>0,~i\in\mathbb{Z}\}$
($\mathbb{Z}$ represents the set of integers), 
$\Phi^{\alpha_0}:=\phi^{i}$ are fields, 
$\Phi^{\alpha_1}$ represents ghost fields, 
and $\Phi^*_{\alpha_l}$ denotes anti-fields of the fields
$\Phi^{\alpha_l}$. 
Hereafter we write a function space created by 
the fields and anti-fields as
$C(\Psi)$. 
It is known that $S_{\rm gh}$ is given by
\begin{equation}
 S_{\rm gh}=S+
\Phi^*_{\alpha_0}R^{\alpha_0}_{\alpha_1}\Phi^{\alpha_1}+
{\cal O}(\Psi^3).
\end{equation}
Note that the anti-fields will be fixed, 
and $\sharp \Psi=\sum_l\sharp \Phi^{\alpha_l}$ 
(see section \ref{sec:PathIntegralTST}).

\begin{figure}[t]
 \begin{center}
  \includegraphics[width=12cm]{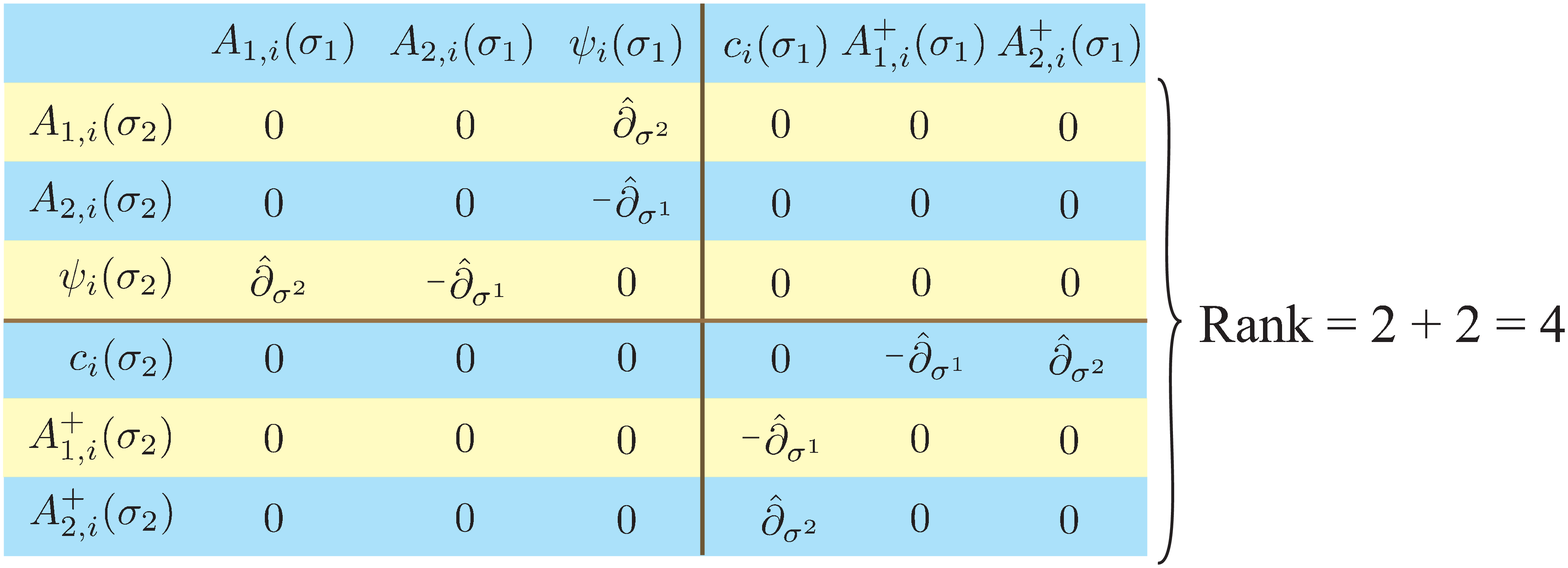}
 \end{center}
 \caption{The Hessian matrix: 
$K[\Psi^\alpha(\sigma_1),\Psi^\beta(\sigma_2)]:=
\frac{\delta}{\delta\Psi^\alpha(\sigma_2)}S
\frac{\delta}{\delta\Psi^\beta(\sigma_1)}$
The first column and first raw represent the right-hand side 
and the left-hand side of variation functions, respectively. 
$\hat{\partial}_{\sigma^j}:=
\delta(\sigma-\sigma_1)
\partial_{\sigma^j}
\delta(\sigma -\sigma_2)$ 
represents a non-trivia Noether current ($j=1,2$). 
The Hessian is block diagonal matrix; 
the ranks of the upper left and the lower right parts are $2=3-1$. 
Therefore, the total rank of the Hessian is $2+2=4$ ($i$ is fixed).
}\label{fig:Hessian}
\end{figure}

For the topological string theory, the fields $\phi^\alpha$ are U(1)
gauge fields $A_{i,\mu}$ and scalar fields $\psi^i$ with $i=1,\dots,N$
and $\mu=1,2$; namely, 
$\Phi^{\alpha_0}\equiv\phi^\alpha:=(A_{i,\mu},\psi^i)$. 
Since $\sharp A_{i,\mu}=2N$ and $\sharp \psi^i=N$, 
the fields number $\sharp \phi^\alpha$ is $3N$. 
The action (\ref{eq:TS}) has the U(1) gauge invariance:
\begin{eqnarray}
 \delta_0A_{\mu,i}&=&\partial_\mu\delta_i^j \chi_j,\\
 \delta_0\psi^i&=&0,\\
 \delta_0\chi_i&=&0
\end{eqnarray}
with $\chi_i$ represents a scalar function ($i=1,\dots,N$). 
Therefore, the topological string theory has $2N$
linear-independent nontrivial symmetry generators. 
Here we replace the scalar fields $\\chi_i$ 
with ghost fields $c_i$ (BRST transformation). 
Moreover, we add antifields $A^*_{i,\mu}$; 
since the gauge transformation does not connect to 
$\psi$ and the other fields, we does not add $\psi^*$
(the space of fields and ghost fields has $2N$ symmetry generators, 
and the space of anti-fields and the anti-ghosts also have 
$2N$ symmetry generators corresponding to U(1) gauge symmetry; 
see Figure~\ref{fig:Hessian}):
\begin{equation}
 R^{(\mu,i)}_\beta=\partial_\mu\delta^i_\beta,\ \ \ \ 
(\beta=1, 2, \dots, N).
\end{equation}
In this case, $\sharp K(\phi^\alpha,\phi^\beta)=3N-2N$; 
on the other hand, the action is a function of 
$3N$ fields $(A_{\mu,i},\psi^i)$, 
$N$ ghost fields $c_i$ and $2N$ anti-fields $A^*_{i,\mu}$. 
Therefore, a rank of the Hessian corresponding to $(S_0)_{\rm gh}$ 
is calculated by
\begin{eqnarray}
 {\rm rank}K(\Psi,\Psi)|_{\Psi_c}&=&{\rm rank}K(\phi,\phi)|_{\Psi_c}
 +\sharp c_i+\sharp A^*_{i,\mu}\nonumber\\
 &=&N+N+2N\nonumber\\
 &=&4N.
\end{eqnarray}
Since $\sharp \Phi=4N$ (antifields will be fixed), 
the field number of the path-integral of $(S_0)_{\rm gh}$ is equal to
the rank of the Hessian of $(S_0)_{\rm gh}$; 
hence, the path-integral of the action $(S_0)_{\rm gh}$ become
well-defined.

Finally, the gauge invariance action is written by
\begin{equation}
 (S_0)_{\rm gh}=S_0+\int_{\Sigma}(A^i)^+\wedge \delta_0A_i
\end{equation}
with $A_i:=A_{\mu,i}d\sigma^\mu$ and 
 $\delta_0A_{\mu,i}=\partial_\mu\delta_i^j c_j$, 
where we define $\Phi^+_\alpha$ using 
a Hodge operator $*_H$: $\Phi^+_\alpha\equiv*_{\rm H}\Phi^*_\alpha$ 
(the definition of the Hodge operator is depend on the geometry
of the disk $\Sigma$; 
we will introduce the explicit definition in 
section~\ref{sec:PerturbationTST}), which is also called as 
the anti-field.

\subsubsection{Condition of gauge invariance of classical
   action}\label{sec:CGC}

In this section, 
we will add interaction terms:
$S_{\rm gh}:=(S_0)_{\rm gh}+g(S_1)_{\rm gh}+\cdots$, 
where $g$ represents an expansion parameter, and 
we will see that $S_{\rm gh}$ is uniquely fixed except 
a certain two form $\alpha$ by a gauge invariance condition. 
Note that $\alpha$ satisfies the Jacobi's identity. 
Therefore, we can identify $\alpha$ with the Poisson bracket. 

First we discuss the gauge invariant condition. 
If we identify the fields and anti-fields with coordinates $q$ and canonical
momentum $p$, i.e., 
$(\Phi^{\alpha_i},\Phi^*_{\alpha_i})\leftrightarrow(q^{\alpha_i},p_{\alpha_i})$, 
and we also identify the action $S$ and the Hamiltonian H: 
$S\leftrightarrow H$. 
In the analytical mechanics, 
$\delta_{\rm am} :=\{H,~\}$ represents a transform along the surface
$H(q,p)={\rm constant}$, i.e., $\delta_{\rm am}$ holds the Hamiltonian. 
Similarly, 
we can define a gauge transformation, 
which holds the action $S$,
using the Poisson bracket in the two-dimension field theory. 
It is known as the Batalin-Vilkovisky (BV) bracket~\cite{Batalin198127,PhysRevD.28.2567};
the definitions of the bracket are  
\begin{equation}
\{f,g\}_{\rm BV}:=\sum_{\substack{ \alpha_i \\ i=0, 1, \ldots}}
\left(\frac{\delta f}{\delta\Phi^{\alpha_i}}
\frac{\delta g}{\delta\Phi^*_{\alpha_i}}-
\frac{\delta f}{\delta\Phi^*_{\alpha_i}}
\frac{\delta g}{\delta\Phi^{\alpha_i}} 
\right)
\end{equation}
with $f,~g\in C(\Psi)$.

The BV bracket has the ghost number $1$, then a BV-BRST operator
$\delta_{\rm BV}:=\{S,~\}_{\rm BV}$ adds one ghost number. 
The BV bracket satisfies the following
equations:
\begin{eqnarray}
&&\{f,g\}_{\rm BV}=-(-1)^{({\rm gh}(f)+1)({\rm gh}(g)+1)}\{g,f\}_{\rm BV},\label{eq:BV1}\\
&&(-1)^{({\rm gh}(f)+1)({\rm gh}(h)+1)}\{f,\{g,h\}_{\rm BV}\}_{\rm BV}+{\rm
 cyclic}=0,\label{eq:BV2}\\ 
&&\{f,gh\}_{\rm BV}=\{f,g\}_{\rm BV}h+(-1)^{({\rm gh}(f)-1){\rm
 gh}(g)}g\{f,h\}_{\rm BV}\label{eq:BV3} 
\end{eqnarray}
with $f,~g,~h\in C(\Psi)$. 

Using the BV-BRST operator, the gauge invariance of action $S$ is
written as $\delta_{\rm BV}S=0$, i.e., 
\begin{equation}
 \{S,S\}_{BV}=0,\label{eq:gaugeinvariance}
\end{equation}
which is called the classical master equation. 
We use this equation and Eqs.~(\ref{eq:BV1}) and (\ref{eq:BV2}); 
we obtain $\delta^2_{\rm BV}=0$, which corresponds to the condition of 
the BRST operator: $\delta^2_{\rm BRST}=0$ ($\delta_{\rm BRST}$ is the
BRST operator). 
Therefore, the DV-BRST operator is the generalized BRST one.

Next, we discuss generalization of the topological field theory. 
Let us write a generalized action $S_{\rm gh}$ as
\begin{equation}
 S_{\rm gh}=(S_0)_{\rm gh}+g(S_1)_{\rm gh}+g^2(S_2)_{\rm gh}+\cdots,
\end{equation} 
where $g$ is an expansion parameter.
Using gauge invariance condition (\ref{eq:gaugeinvariance}), 
$(S_n)_{\rm gh}$ ($n=1,2,\cdots$) is given by 
a solution of the following equation:
\begin{equation}
 \left.\frac{\partial^n}{\partial g^n}\{S_{\rm gh},S_{\rm gh}\}_{\rm BV}\right|_{g=0}=0.
\end{equation}
The general solution is given by~\cite{Izawa}
\begin{eqnarray}
 (S_1)_{\rm gh}&=&\int_\Sigma d^2\sigma\Bigg[\frac{1}{2}\alpha^{ij}(A_iA_j-2\psi^+_ic_j)+
\frac{\partial\alpha^{ij}}{\partial\psi^k}\left(\frac{1}{2}(c^+)^kc_ic_j
-(A^+)^kA_ic_j\right)\nonumber\\
&&+\frac{1}{4}\frac{\partial^2\alpha^{ij}}
{\partial\psi^k\partial\psi^l}(A^+)^k(A^+)^lc_ic_j\Bigg],\\
 (S_{n>1})_{\rm gh}&=&0
\end{eqnarray}
with $(A^+)^i\equiv*_{\rm H}A_{\mu,i}^*=d\sigma^\mu\varepsilon_{\mu\nu}A_{\nu,i}^*$, 
$\psi^+_i\equiv*_{\rm H}\psi_i^*=\varepsilon_{\mu\nu}d\sigma^\mu \wedge d\sigma^\nu\psi^*_i$ 
and 
$(c^+)^i\equiv*_{\rm H}(c^*)^i=\varepsilon_{\mu\nu}d\sigma^\mu\wedge
d\sigma^\nu(c^*)^i$ ($\varepsilon_{\mu\nu}=-\varepsilon_{\nu\mu},~\varepsilon_{12}=1$),
where $\alpha^{ij}$ is a function of $\psi$, and satisfies the following
equation:
\begin{equation}
 \frac{\partial\alpha^{ij}}{\partial\psi^m}\alpha^{mk}+
 \frac{\partial\alpha^{jk}}{\partial\psi^m}\alpha^{mi}+
 \frac{\partial\alpha^{kl}}{\partial\psi^m}\alpha^{mj}=0.
\end{equation}
Here, if we identify $\psi^i$ with $x^i$, this equation is 
the Jacobi identity of Poisson bracket. 
Therefore, we can identify the Poisson bracket with the topological
string theory.

\subsubsection{Gauge invariance in path integral}\label{sec:PathIntegralTST}
Now we discuss the path integral of the topological string theory 
$\int {\cal D}\Phi V(\Psi)$
with $V(\Psi)={\cal O}e^{\frac{i}{\hbar}S}$, 
and an observable quantity operator ${\cal O}$. 
Note that this path-integral does not include integrals in terms of the
anti-fields. 
Therefore, we must fix the anti-fields; 
then, we consider that the anti-field $\Phi^*$ is a function of the
field $\Phi$, i.e., $\Phi^*=\Omega(\Phi)$ and 
$\Omega\in {\cal C}(\Phi=\Psi)$ 
Namely, the path integral is defined by
\begin{equation}
 \left.\int{\cal D}\Phi V(\Psi)\right|_{\Phi^*=\Omega}.
\end{equation}

A choice of $\Omega(\Phi)$ is corresponding to the gauge fixing in the
gauge theory. 
The path integral must be independent to the gauge choice (gauge invariance). 
To obtain a gauge invariant condition, 
we take the variation of the path integral in terms of anti-fields, 
and obtain the following gauge invariant condition~\cite{Batalin1977309}:
\begin{equation}
 \triangle_{\rm SD}V(\Psi)=0,\label{eq:QME}
\end{equation}
where we have introduced the Schwinger-Dyson (SD) operator:
\begin{equation}
 \triangle_{\rm SD}:=\sum_{\alpha_l}(-1)^{\alpha_l}
\frac{\delta}{\delta\Phi^{\alpha_l}}\frac{\delta}{\delta\Phi^*_{\alpha_l}},
\end{equation}
where $(-1)^{\alpha_l}$ is defined as follows: 
if $\Phi^{\alpha_l}$ represents a boson, 
$(-1)^{\alpha_l}=(-1)^{{\rm gh}(\Phi^{\alpha_l})}$; 
if $\Phi^{\alpha_l}$ represents a fermion, 
$(-1)^{\alpha_l}=(-1)^{({\rm gh}(\Phi^{\alpha_l})+1)}$.
Equation (\ref{eq:QME}) is called the quantum master equation. 
It is known that the following two conditions are equivalence:
\begin{equation}
 \triangle_{\rm SD}V(\Psi)=0\Longleftrightarrow \Omega=
 \frac{\overrightarrow{\delta}\varphi}{\delta\Phi^a},
 \ \ \ \exists\varphi,\label{eq:gaugefix}
\end{equation}
where $\varphi$ is called the gauge-fixing fermion
(an example will be shown later). 

To perform the path integral, we generalize the classical action
$S_{\rm gh}$ to a quantum action 
$W=S_{\rm gh}+{\rm i}\hbar W_1+({\rm i}\hbar)^2W_2+\cdots$. 
The correction terms $W_n$ ($n=1, 2, \ldots$) are calculated from the
master equation:
\begin{equation}
 \triangle_{\rm SD}e^{\frac{i}{\hbar}W}=0,
\end{equation}
or
\begin{eqnarray}
&& \{S_{\rm gh},S_{\rm gh}\}_{\rm BV}=0,\\
&& \{W_1,S_{\rm gh}\}_{\rm BV}+{\rm i}\hbar\triangle_{\rm SD}S_{\rm gh}=0,\\
&& \{W_2,S_{\rm gh}\}_{\rm BV}+
{\rm i}\hbar\triangle_{\rm SD}W_1+
\frac{1}{2}\{W_1,W_1\}_{\rm BV}=0,\\
&&\ \ \ \ \ \ \ \ \ \ \ \ \ \ \ \ \ \ \ \ \ \cdots\nonumber
\end{eqnarray}
In the case where $\triangle_{\rm SD}S_{\rm gh}=0$, we can put
$W_1=W_2=\cdots=0$. 
Fortunately, the topological string theory satisfies 
$\triangle_{\rm SD}S_{\rm gh}=0$. 
Therefore, we do not have to be concerned about the quantum correction
of the action.  

Finally, we consider the gauge fixing. 
Here we employ the Lorentz gauge:
\begin{equation}
 d*_{\rm H}A_i=0,
\end{equation}
and we add the integral of the Lorentz gauge to $S_{\rm gh}$. 
However, the path integral should hold gauge invariance, i.e., 
the path integral should be independent of gauge fixing term. 
Then, the gauge fixing can be written gauge-fixed fermion: 
\begin{equation}
 \varphi:=\int_\Sigma\gamma^i(d*_{\rm H}A_i)=
-\int_\Sigma d\gamma^i*_{\rm H}A_i,
\end{equation}
where we introduced $N$ fields $\gamma_i$
($i=1, 2, \ldots, N$), and anti-fields $\gamma_i^+$ are given by
\begin{equation}
 \gamma^+_i=\frac{\overrightarrow{\partial}\varphi}{\partial\gamma_i}
=d*_{\rm H}A_i.
\end{equation} 
Now, we employ the Lagrange multiplier method, 
and introduce $N$ scalar fields $\lambda_i$. 
The gauge-fixed action is written by
\begin{eqnarray}
 S_{\rm gf}&=&S_{\rm gh}-\int_\Sigma\gamma^id*_{\rm H}A_i\\
&=&S_{\rm gh}-\int_\Sigma \lambda^i\gamma_i^+.
\end{eqnarray}

The other anti-fields are also fixed by this gauge-fixing fermion:
\begin{eqnarray}
&& \psi^+_i=c^+_i=\lambda^+_i=0,\\
&& A^+_i=*_{\rm H}d\gamma^i.
\end{eqnarray}
Gauge fixed action $S_{\rm gf}$ is written by
\begin{eqnarray}
 S_{\rm gf}&=&\int_\Sigma \Bigg[A_i\wedge d\psi^i
+\frac{1}{2}\alpha^{ij}A_i\wedge A_j
-*_{\rm H}d\gamma^i\wedge\left(
dc_i+\frac{\partial\alpha^{kl}}{\partial\psi^i}A_kc_l\right)\nonumber\\
&&-\frac{1}{4}*_{\rm H}d\gamma^i\wedge*_{\rm H}d\gamma^j
\frac{\partial^2\alpha^{kl}}{\partial\psi^i\partial\psi^j}c_kc_l
-\lambda^id*_{\rm H}A_i\Bigg].
\end{eqnarray}
Here we perform the following variable transformations:
\begin{eqnarray}
 X^i&:=&\psi^i+\theta^\mu A^*_\mu
-\frac{1}{2}\theta^\mu\theta^\nu c^{+i}_\mu\nu,\\
 \eta_i&:=&c_i+\theta^\mu A_{i,\mu}+\frac{1}{2}\theta^\mu\theta^\nu 
\psi^+_{i,\mu\nu},
\end{eqnarray}
where $\theta^\mu\vartheta\nu=-\theta\nu\theta\mu$; 
${\rm gh}(\theta^\mu)=1$. 
For any scalar field $f(u)$ ($u\in\Sigma$), 
$\tilde{f}(u,\theta):=
f(u)+\theta^\mu f^{(1)}_\mu(u)
+\frac{1}{2}\theta^\mu\theta^\nu f^{(2)}_{\mu\nu}$ is 
called as the super field, where 
$f^{(1)}$ and $f^{(2)}$ represent a one-form field and a two-form field,
respectively. 

By using the super fields, the gauge fixed action $S_{\rm gf}$ can be
rewritten as 
\begin{equation}
 S_{\rm gf}=\int_\Sigma\int d^2\theta \left[
\eta_iDX^i-\lambda^id*_{\rm H}A_i+
\frac{1}{2}\alpha^{ij}(X)\eta_i\eta_j
\right],
\end{equation}
where $D:=\theta^\mu\frac{\partial}{\partial u^\mu}$. 
This is the final result in this section. 
Hereafter, we write $S^0_{\rm gf}:=\int_\Sigma \left[\eta_iDX^i-
\lambda^id*_{\rm H}A_i\right]$
and $S^1_{\rm gf}:=\int_\Sigma \alpha^{ij}\eta_i\eta_j/2$. 

\subsection{Equivalence between deformation quantization and topological
  string theory}\label{sec:EDQTST}
We return to the discussion about the deformation quantization. 
Here we see that the equivalence of the deformation quantization and the
topological string theory, and introduce the perturbation theory of the
topological string theory, 
which is equal to Kontsevich's deformation quantization~\cite{Kontsevich}. 

\subsubsection{Path integral as $L_\infty$ map}\label{sec:PathL}
Here we summarize correspondence between Path integral with $L_\infty$
map. 

First we note that the map: 
$\alpha:=\alpha(x)^{\mu\nu}\eta_\mu\eta_\nu/2\mapsto
S_\alpha:=S^1_{\rm gf}=
\int_\Sigma\alpha(x)^{\mu\nu}\eta_\mu\eta_\nu/2$
is isomorphic, because
$\{S_{\alpha_1},S_{\alpha_2}\}_{\rm BV}=S_{\{\alpha_1,\alpha_2\}_{\rm BV}}$. 

SD operator satisfies the conditions of codifferential operator $Q$ in
$L_{\infty}$ algebra, where the vector space and the degree of the space
correspond to ${\cal C}(\Psi)$ and the ghost number, respectively. 

The path integral $\int e^{\frac{i}{\hbar}S^0_{\rm gf}}$ gives the
deformation quantization ${\mathscr F}_0+{\mathscr F}^1$. 
The master equation 
\begin{eqnarray}
 Q e^{\frac{i}{\hbar}S_\alpha}=0
\end{eqnarray}
with $Q=\triangle_{\rm SD}$ is corresponding to the $L_\infty$ map's 
condition $Q{\mathscr F}=0$

For 
$\alpha_r:=\alpha^{i_1,i_2,\cdots,i_m}\eta_{i_1}\eta_{i_2}\cdots
\eta_{i_m}/m!$ with a positive integer $m$, 
${\mathscr F}_n:V_1^{\otimes n}\otimes\rightarrow V_2$ is given by
\begin{equation}
 {\mathscr F}_n(\alpha_1, \ldots, \alpha_n)
(f_1\otimes\cdots\otimes f_m)(x)
:=\int e^{\frac{i}{\hbar}S^0}\frac{i}{\hbar}S_{\alpha_1}\cdots\frac{i}{\hbar}
 S_{\alpha_n}{\cal O}(f_1, \ldots, f_m),
\end{equation}
where $S_{\alpha_r}$ is the expansion of $S_\alpha$, 
and is defined as
\begin{equation}
 S_{\alpha_r}:=\left.\left(
\int_\Sigma\frac{1}{m!}\alpha^{i_1\cdots i_m}(X)
\eta_{i_1}\cdots\eta_{i_m}\right)\right|_{\Phi^*=\partial\varphi},
\end{equation} 
and ${\cal O}$ is chosen to satisfy 
\begin{equation}
 Q\circ{\mathscr F}={\mathscr F}\circ Q,
\end{equation}
where ${\mathscr F}:={\mathscr F}_0+{\mathscr F}^1
+{\mathscr F}^2+\cdots$. 
We put ${\cal O}$ as follow: 
\begin{eqnarray}
 {\cal O}(f_1, \ldots, f_m)&=&\int_{B_m}[X(t_1,\theta_1))
\cdots f_m(X(t_m,\theta_m))]_{(m-2)}\delta_x(X(\infty)),\\
&\equiv&
\int_{1=t_1>t_2>\cdots>t_m=0}
f_1(\psi(1))\prod_{k=2}^{m-1}
\partial_{i_k}\left[
f(\psi(t_{k}))A^{+i_k}(t_k)
\right]
f_m(\psi(0))\delta_x(\psi(\infty))
\end{eqnarray}
where the subscript ${(m-2)}$ denotes
that $(m-2)$ forms are picked up from the products of super fields, 
and $B_m$ represents the surface of the disk $\Sigma$, i.e., 
$t$ is the parameter specifying the position on the boundary 
$\partial\Sigma$ ($1=t_1>t_2>\cdots>t_{m-1}>t_m=0)$. 

To be exact, the action and fields include gauge fixing terms, 
ghost fields and anti-field. 
Finally, the deformation quantization is given as follow:
\begin{equation}
 (f\star g)(x)=\int{\cal D}\Phi{f(\psi(1))g(\psi(0))
\delta(x^i-\psi^i(\infty))}e^{\frac{i}{\hbar}S_{\rm gf}}.
\end{equation}

\subsubsection{Perturbation theory}\label{sec:PerturbationTST}
Now we see that the perturbation theory of the topological string
theory. 
First, we write the action as $S_{\rm gf}=S^0_{\rm gf}+S^1_{\rm gf}$. 
The first term is defined as
\begin{equation}
 S^0_{\rm gf}=\int_\Sigma \left[A_i\wedge(d\xi^i+
*_{\rm H}d\lambda^i)+c_id*_{\rm H}d\gamma^i\right],
\label{eq:S0}
\end{equation}
where $\xi^i\equiv\psi^i- x^i$, 
and we have expanded $\psi^i$ around $x^i$. 
The path integral of an observable quantity $\langle{\cal O}\rangle$ is given by
\begin{equation}
 \int e^{\frac{i}{\hbar}S_{\rm gf}}
{\cal O}=\sum_{n=0}^\infty\frac{i^n}{\hbar^nn!}\int
  e^{\frac{i}{\hbar}S^0_{\rm gf}}(S^1_{\rm gf})^n{\cal O},
\end{equation}
where 
$\int:=\int{\cal D}\xi {\cal D}A{\cal D}
c{\cal D}\gamma {\cal D}\lambda$. 
This expansion corresponds to the summation of all diagrams by the
contractions of all pairs in terms of fields and
ghost fields. 
From equation~(\ref{eq:S0}), propagators are inverses of
\begin{equation}
 d\oplus*_{\rm H}d,\ \ \ \ d*_{\rm H}d.
\end{equation}
Here we assume that the disk is the upper complex plane: 
$\Sigma=\{z~|~z=u+iv,~u,v\in\mathbb{R},~v\geq 0\}$ with
$i^2=-1$, 
and the boundary is $\partial\Sigma=\{z~|~z=u,~u\in\mathbb{R}\}$. 
($\mathbb{R}$ represents the real number space, and $z$ denotes a complex number.)
The Hodge operator $*_{\rm H}$ is defined by
\begin{eqnarray}
\left\{
\begin{array}{c}
 *_{\rm H}du=dv\\
 *_{\rm H}dv=-du
\end{array}\right.
\longmapsto 
\left\{
\begin{array}{c}
 *_{\rm H}dz=-idz\\
 *_{\rm H}d\overline{z}=id\overline{z}
\end{array}
\right.,
\end{eqnarray}
where $\overline{z}$ represents the complex conjugate of $z$. 
Moreover,
\begin{eqnarray}
 &&d_z=du\frac{\partial}{\partial u}
+dv\frac{\partial}{\partial v}
=dz\frac{\partial}{\partial z}
+d\overline{z}\frac{\partial}{\partial\overline{z}},\\
&&\delta_z(w):=\delta(w-z)du_w\wedge dv_w,~\int\delta_z(w)=1,
\end{eqnarray}
where $w\in\mathbb{C}$ with the complex number plane $\mathbb{C}$, 
and $w\equiv u_w+iv_w$. 

Now, we calculate Green functions of $d\oplus *_{\rm H}d$ and $d*_{\rm H}d$, 
because the Green functions are inverses of these operators:
\begin{equation}
 D_wG(z,w)={\rm i}\hbar\delta_z(w),
\end{equation}
where $D_w=d_w\oplus *_{\rm H}d_w~{\rm or}~d_w*_{\rm H}d_w$. 
The solution depends on the boundary condition. 
In the case that $z$ and $w$ satisfy the Neumann boundary condition, 
a solution is a function of
\begin{equation}
 \phi^h(z,w):=\frac{1}{2i}{\rm log}
\frac{(z-w)(z-\overline{w})}{(\overline{z}-w)(\overline{z}-\overline{w})}.
\end{equation}
On the other hand, $z$ and $w$ satisfy the Dirichlet boundary condition,
a solution is a function of
\begin{equation}
 \psi^h(z,w):=\log\left|\frac{z-w}{z-\overline{w}}\right|.
\end{equation}
The Neumann boundary condition is $0=\partial_{u^1}G(z,w)|_{u^2=0}$, 
and the Dirichlet boundary condition is
$0=\partial_{u^2}G(z,w)|_{u^1=0}$. 

The propagators are given by
\begin{eqnarray}
 \langle\gamma^k(w)c_j(z)\rangle&=&\frac{{\rm i}\hbar}{2\pi}\delta^k_j\psi^h(z,w),\\
 \langle\xi^k(w)A_j(z)\rangle&=&\frac{{\rm i}\hbar}{2\pi}\delta^k_jd_z\phi^h(z,w),\\
 \langle(*_{\rm H}d\gamma^k)(w)c_j(z)\rangle&=&
\frac{{\rm i}\hbar}{2\pi}\delta^k_j\delta_w\phi^h(z,w),
\end{eqnarray}
and so on. 
From these propagators, we can obtain diagram rules corresponding to the
deformation quantization. 
In section \ref{sec:DiagramRules}, we will introduce exact diagram
rules. 

To obtain the star product, we choice
\begin{equation}
 {\cal O}_x=f(X(1))g(X(0))\delta_x(\psi(\infty))
\end{equation}

\subsection{Diagram rules of deformation quantization}\label{sec:DiagramRules}
From the perturbation theory of the topological string theory, 
we can obtain the following diagram rules of the star product, 
which is first given by Kontsevich~\cite{Kontsevich,Sugimoto}:
\begin{eqnarray}
(f\star g)(x)=f(x)g(x)+\sum^\infty _{n=1}\left(\frac{{\rm i}\hbar }{2}\right)^n
\sum_{\Gamma \in G_n}w_{\Gamma }B_{\Gamma ,\alpha }(f,g)\label{K*}.
\end{eqnarray}
where $\Gamma $, $B_{\Gamma ,\alpha }(f,g)$ and $w_{\Gamma }$ are
defined as follows:
\begin{Definition}
$G_n$ is a set of the graphs $\Gamma $ 
which have $n+2$ vertices and $2n$ edges. 
Vertices are labeled by symbols 
``$1$'', ``$2$'', $\ldots$ , ``$n$'', ``$L$'', and ``$R$''. 
Edges are labeled by symbol $(k,v)$, 
where $k=1, 2, \ldots , n$, $v=1, 2, \ldots , n, L, R$, and $k\not =v$. 
$(k,v)$ represents the edge which 
starts at ``$k$'' and ends at ``$v$''. 
There are two edges starting from each vertex with 
$k=1, 2, \ldots , n$; 
$L$ and $R$ are the exception, i.e., 
they act only as the end points of the edges. 
Hereafter, $V_{\Gamma }$ and $E_{\Gamma }$ represent the set of 
the vertices and the edges, respectively. 
\end{Definition}
\begin{Definition}
$B_{\Gamma ,\alpha }(f,g)$ is the operator defined by:
\begin{eqnarray}
B_{\Gamma ,\alpha }(f,g)&:=&
\sum_{I:E_{\Gamma }\rightarrow \{i_1,i_2,\cdots,i_{2n}\}} 
\left [\prod ^n_{k=1}\left (\prod _{e\in E_{\Gamma },
e=(k,*)}\partial _{I(e)}\right )
\alpha ^{I((k,v_k^1),(k,v_k^2))}\right]\times \nonumber\\ 
&&
\left[\left(
\prod _{e\in E_{\Gamma },e=(*,L)}\partial _{I(e)}
\right)f\right]\times 
\left[\left (\prod _{e\in E_{\Gamma },
e=(*,R)}\partial _{I(e)}
\right)g\right],
\end{eqnarray} 
where, $I$ is a map from the list of edges 
$((k,v_k^{1,2})), k=1,2,\ldots,n$ to integer numbers 
$\{i_1,i_2,\cdots,i_{2n}\}$. 
Here $1\leq i_n\leq d$; $d$ represents a dimension of the manifold $M$. 
$B_{\Gamma ,\alpha }(f,g)$ corresponds 
to the graph $\Gamma $ in the following way: 
The vertices ``$1$'', ``$2$'', $\ldots$ , ``$n$'', 
correspond to the Poisson structure $\alpha ^{ij}$. 
$R$ and $L$ correspond to the functions $f$ and $g$, respectively. 
The edge $e=(k,v)$ represents the differential operator
$\partial_{(i~{\rm or}~j)}$ acting on the vertex $v$. 

The simplest diagram for $n=1$ is shown in Fig.~$\ref{example} (a)$, 
which corresponds to the Poisson bracket: 
$\{f,g\}=
\sum_{i_1,i_2}\alpha ^{i_1i_2}
(\partial_{x^{i_1}}f)(\partial _{x^{i_2}}g)$. 
The higher order terms are the generalizations of this Poisson bracket. 

Figure~$\ref{example} (b)$ shows 
a graph $\Gamma_{{\rm ex.2}}$ with $n=2$ corresponding to the list of edges 
\begin{eqnarray}
((1,L),(1,R),(2,R),(2,3));
\end{eqnarray}
\begin{figure}[t]
\begin{center}
\includegraphics[width=6cm,clip]{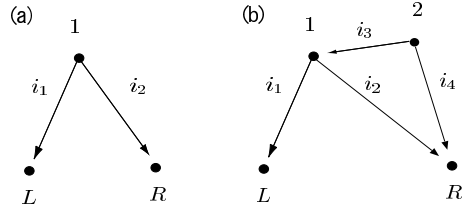}
\end{center}
\caption{
(a): The graph $\Gamma_{{\rm ex.1}}\in G_1$ 
corresponding to Poisson bracket. 
(b): A graph $\Gamma \in G_2$ correspond to the list of edges: 
$((1,L),(1,R),(2,R),(2,1))\mapsto \{i_1,i_2,i_3,i_4\}$.  
}
\label{example}
\end{figure}
in addition, the operator $B_{\Gamma_{{\rm ex}.2},\alpha }$ is given by
\begin{eqnarray}
(f,g)\mapsto \sum _{i_1,\cdots ,i_4}(\partial _{x^{i_3}}
\alpha ^{i_1i_2})\alpha ^{i_3i_4}
(\partial _{x^{i_1}}f)(\partial _{x^{i_2}}\partial _{x^{i_4}}g). 
\end{eqnarray}

\end{Definition}
\begin{Definition}
We put the coordinates for the vertices 
in the upper-half complex plane 
$H_+:=\{z\in {\mathbb C}~|~{\rm Im}(z)>0\}$ 
$(\mathbb{C}$ represents the complex plain; 
${\rm Im}(z)$ denotes the imaginary part of $z)$. 
Therefore, $R$ and $L$ are put at $0$ and $1$, respectively. 
We associate a weight $w_{\Gamma }$ with each graph $\Gamma \in G_n$ as
\begin{eqnarray}
w_{\Gamma }:=\frac{1}{n!(2\pi)^{2n}}\int_{{\cal H}_n}\bigwedge_{k=1}^n
\left (d\phi ^h_{(k,v_k^1)}\wedge d\phi ^h_{(k,v_k^2)}\right ),
\end{eqnarray}
where $\phi $ is defined by 
\begin{eqnarray}
\phi ^h_{(k,v)}:=\frac{1}{2i}{\rm Log}
\left (\frac{(q-p)(\bar q-p)}{(q-\bar p)(\bar q-\bar p)}\right ).
\end{eqnarray}
$p$ and $q$ are the coordinates of the vertexes ``$k$'' and ``$v$'', respectively. 
$\bar p$ represents the complex conjugate of $p\in\mathbb{C}$. 
${\cal H}_n$ denotes the space of configurations of 
$n$ numbered pair-wise distinct points on $H_+$: 
\begin{eqnarray}
{\cal H}_n:=\{ (p_1,\cdots ,p_n)~|~p_k\in H_+,~p_k\not =p_l~\text{for}~k\not =l\}.
\end{eqnarray}
Here we assume that $H_+$ has the metric:
\begin{equation}
 ds^2=(d({\rm Re}(p))^2+d({\rm Im}(p))^2)/({\rm Im}(p))^2,\label{eq:Metric}
\end{equation} 
with $p\in H_+$; 
$\phi ^h(p,q)$ is the angle which is defined by 
$(p,q)$ and $(\infty,p)$, i.e., $\phi ^h(p,q)=\angle pq\infty $ 
with the metric~$(\ref{eq:Metric})$. 
For example, 
$w_{\Gamma_{{\rm ex}.1}}$ corresponding to Fig.~$\ref{example} (a)$ is calculated as:
\begin{equation}
w_{\Gamma_{{\rm ex}.1}}=\frac{2}{1!(2\pi )^2}\int _{{\cal H}_1}
d\frac{1}{2i}{\rm Log}\left (
\frac{p^2}{\overline p^2}\right )\wedge
d\frac{1}{2i}{\rm Log}\left (
\frac{(1-p)^2}{(1-\overline p)^2}\right ) =1,\label{w1}
\end{equation}
where we have included the factor ``$2$'' arising from 
the interchange between two edges in $\Gamma$. 
$w_{\Gamma }$ corresponding to the Fig.~$\ref{example} (b)$ is
\begin{eqnarray}
w_{\Gamma_{(b)}}&=&\frac{1}{2!(2\pi )^4}\int _{{\cal H}_2}
d\frac{1}{i}{\rm Log}
\left(\frac{p_1}{\overline p_1}\right)\wedge
d\frac{1}{i}{\rm Log}
\left(\frac{1-p_1}{1-\overline{p}_1}
\right)\wedge
d\frac{1}{i}{\rm Log}
\left(\frac{p_2}{\overline{p}_2}
\right)\wedge 
d\frac{1}{2i}{\rm Log}
\left(\frac{(p_1-p_2)(\overline{p}_1-p_2)}
{(p_1-\overline{p}_2)(\overline{p}_1-\overline{p}_2)}
\right)
\nonumber\\
&=&\frac{1}{2!(2\pi )^4}\int _{{\cal H}_2}
d\frac{1}{i}{\rm Log}
\left (\frac{p_1}{\overline p_1}\right )\wedge
d\frac{1}{i}{\rm Log}
\left (\frac{1-p_1}{1-\overline{p}_1}
\right)\nonumber\\
&&\wedge
d\left (2\arg (p_2)\right )\wedge 
d|p_2|\frac{\partial }{\partial |p_2|}
\frac{1}{2i}{\rm Log}\left (
\frac{(p_1-p_2)(\overline{p}_1-p_2)}
{(p_1-\overline{p}_2)(\overline{p}_1-\overline{p}_2)}
\right)\nonumber\\
&=&\frac{1}{2!(2\pi )^4}\int _{{\cal H}_2}
d\frac{1}{i}{\rm Log}
\left(\frac{p_1}{\overline{p}_1}
\right)\wedge
d\frac{1}{i}{\rm Log}
\left(\frac{1-p_1}{1-\overline{p}_1}
\right)\wedge
d\frac{1}{i}{\rm Log}
\left(\frac{p_2}{\overline{p}_2}
\right)\wedge 
d\frac{1}{i}{\rm Log}
\left(\frac{1-p_2}{1-\overline{p}_2}
\right)\nonumber\\
&=&\frac{w_1^2}{2!}\nonumber\\
&=&\frac{1}{2},\label{w2}
\end{eqnarray}
where $p_1$ and $p_2$ are the coordinates of 
vertexes ``$1$'' and ``$2$'', respectively. 
Here, we have used the following facts:
\begin{eqnarray}
\int _0^\infty d|p_2|\partial _{|p_2|}{\rm Log}\left (
\frac{(p_1-p_2)(\overline{p}_1-p_2)}
{(\overline{p}_1-p_2)(\overline{p}_1-\overline{p}_2)} 
\right)&=&
\lim_{\Lambda \to \infty }{\rm Log}
\left(
\frac{(p_1-\Lambda e^{i\arg (p_2)})
(\overline{p}_1-\Lambda e^{i\arg(p_2)})}
{(p_1-\Lambda e^{-i\arg(p_2)})
(\overline{p}_1-\Lambda e^{-i\arg(p_2)})}
\right)\nonumber\\ 
&=&\lim _{\Lambda \to \infty }{\rm Log}\left (
\frac{(1-\Lambda e^{i\arg (p_2)})(1-\Lambda e^{i\arg (p_2)})}{(1-\Lambda
e^{-i\arg (p_2)})(1-\Lambda e^{-i\arg (p_2)})}\right ),\nonumber\\ 
\int _{|p_1|> \Lambda }
d{\rm Log}\left (\frac{p_1}{\overline p_1}\right )\wedge
d{\rm Log}\left (\frac{1-p_1}{1-\overline p_1}\right )
&\stackrel{\Lambda \rightarrow \infty }{\longrightarrow }&
\int _{|p_1|> \Lambda }d{\rm Log}\left (\frac{p_1}{\overline p_1}\right )\wedge 
d{\rm Log}\left (\frac{p_1}{\overline p_1}\right )=0.
\end{eqnarray}

Generally speaking, the integrals are entangled for $n \geq 3$ graphs,
 and the weight of these are not so easy to evaluate as Eq.~$(\ref{w2})$. 
\end{Definition}

Note that the above diagram rules also define the twisted element as the
following relation: $(f\star g)\equiv\mu\circ {\cal F}(f\otimes g)$. 

\subsection{Gauge invariant star product}\label{sec:TwistedElement}
From Eq.~(\ref{eq:SU(2)Poisson}), the Poisson structure
corresponding to our model is 
\begin{eqnarray}
&&\alpha^{ij}=\left(
\begin{array}{ccc}
0&\eta ^{\mu \nu }&0\\
-\eta ^{\mu \nu
 }&-q\hat{F}^{\mu\nu}&-q\epsilon^{abc}s^aA^b_\mu\\ 
0&q\epsilon^{abc}s^aA^b_\mu&\epsilon^{abc}s^c
\end{array}
\right),
\end{eqnarray}
where the symbols $i$ and $j$ represent indexes of the phase space
$(T\bm{X},\omega\bm{p},\bm{s})$. 
We separate the Poisson structure as follows:
\begin{eqnarray}
\alpha&:=&
\left(
\begin{array}{ccc}
0&\eta ^{\mu\nu}&0\\
-\eta ^{\mu \nu }&0&0\\
0&0&0
\end{array}
\right)+\left(
\begin{array}{ccc}
0&0&0\\
0&0&-q\epsilon^{abc}A^b_\mu s^a\\
0&q\epsilon^{abc}s^aA^b_\mu&\epsilon^{abc}s^a
\end{array}
\right)+\left(
\begin{array}{ccc}
0&0&0\\
0&q\hat{F}^{\mu\nu}&0\\
0&0&0
\end{array}
\right)
\nonumber\\
&\equiv &\alpha_0+\alpha_A+\alpha_F.\label{P}
\end{eqnarray}
Here, for $f=f_0+f_a\sigma^a$, 
$\partial_{s^a}f:=f_a$ ($a=x,y,z$), 
where $f_{0,x,y,z}$ are functions $X$ and $p$. 
Because $\alpha_0$ is constant and $\alpha_A$ and $\alpha_F$ are
functions of $X^\mu$ and $\bm{s}$, 
and any function $f$ is written as $f=f_0+\sum_{a=x,y,z}f_as^a$ 
($f_{0,a}$ only depends on $X$ and $p$), 
then we obtain additional diagram rules:
\begin{list}{}{}
\item[A1.] Two edges starting from $\alpha _F$ connect with both
	   vertices ``$L$'' and ``$R$''. 
\item[A2.] At least one edge from vertices $\alpha_0$ or $\alpha_F$
	   connect with vertices ``$L$'' or ``$R$''. 
\item[A3.] A number of the edges entering $\alpha_A$ is 
	   one or zero. 
\end{list}

We also separate the graph $\Gamma $ into 
$\Gamma _{\alpha _0}$, $\Gamma_{\alpha _A}$ and $\Gamma _{\alpha _F}$. 
Here, we define the numbers of vertices 
$\alpha_0$, $\alpha_A$, and $\alpha_F$ as
$n_{\alpha_0}$, $n_{\alpha_A}$, and $n_{\alpha_F}$, respectively. 
$\Gamma _{\alpha _F}$ is the graph consisted by vertices corresponding
to $\alpha _F$, and ``$L$'' and ``$R$'', and edges starting from 
these vertices. 
We consider $\Gamma _{\alpha _F}$ as a cluster, and define 
$\Gamma_{\alpha _A}$ as the graph consisted by the vertices corresponding to
$\alpha_A$, which acts on the cluster corresponding to
$\Gamma_{\alpha_F}$. 
$\Gamma_{\alpha_0}$ is the rest of the graph 
$\Gamma $ without $\Gamma_{\alpha_A}$ and $\Gamma _{\alpha _F}$. 
Here, we label vertexes $\Gamma_{\alpha_F}$, $\Gamma_{\alpha_A}$ and $\Gamma
_{\alpha _0}$ by ``$k=1-n_{\alpha_F}$'',
``$k=(n_{\alpha_F}+1)-(n_{\alpha_F}+n_{\alpha_A})$'' and
``$k=(n_{\alpha_F}+n_{\alpha_A}+1)-(n_{\alpha_F}+n_{\alpha_A}+n_{\alpha_0})$'',
respectively. 
The edge starting from ``$k$'' and ending to ``$v_k^{1,2}$'' represents
$(k,v_k^{1,2})$. 

Next, we calculate weight $w_{n_{\alpha _F}}$ and the operator
$B_{\Gamma _{\alpha _F},\alpha _F}$ corresponding to $\Gamma _{\alpha
_F}$, and later those for $\Gamma _{\alpha _{A or 0}}$. 

\vskip 0.5cm
\noindent
\textit{Separation of graph $\Gamma $}

We now sketch the proof of 
$w_{\Gamma }B_{\Gamma ,\alpha }=w_{n _{\alpha
_0}}B_{\Gamma _{\alpha _0},\alpha _0}\cdot
w_{n_{\alpha_A}}B_{\Gamma_{\alpha_A},\alpha_A}\cdot w_{n
_{\alpha _F}}B_{\Gamma _{\alpha _F},\alpha _F}$, 
where $w_{n _{\alpha_a}}=
\frac{w_1^{n_{\alpha_a}}}{n_{\alpha_a}!}$
for $a=0, A, F$, and $w_1$ is given by Eq.~(\ref{w1}). 
\begin{figure}[t]
\begin{center}
\includegraphics[width=3cm,clip]{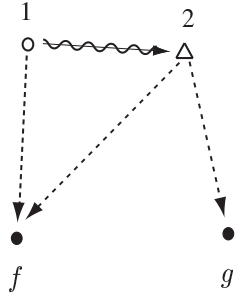}
\end{center}
\caption{A four vertexes graph, where the white circle and the white
 square represent $\alpha _A$ and $\alpha _F$, respectively; the
 dotted arrow, waved arrow, and real arrow represent 
 $\partial_p$, $\partial_s$, and $\partial_X$, respectively.  
}
\label{graph0.eps}
\end{figure}

\begin{figure}[t]
\begin{center}
\includegraphics[width=12cm,clip]{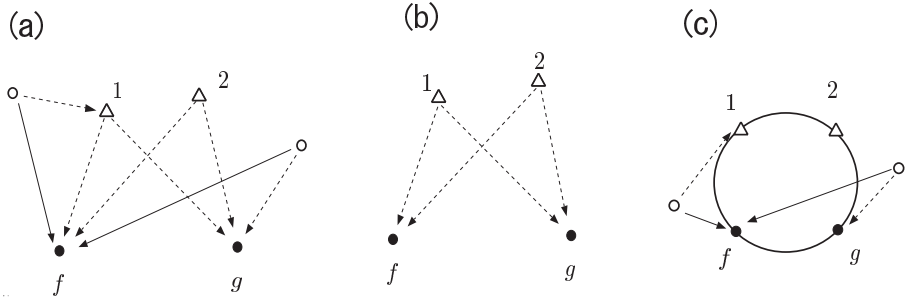}
\end{center}
\caption{
This figure shows the calculation method of the graph $(a)$, where the
 dotted arrow and real arrow represent the derivative with respect to 
$p$ and $X$, respectively, and the white circle and the white triangle
 represent $\alpha _0$ and $\alpha _F$, respectively. 
We rewrite the graph $(a)$ as the graph $(c)$ which is given by 
the cluster represented by the big circle and the operators into it, 
where the big circle represents the graph $(b)$. 
}\label{GD2}
\end{figure}

From the additional rule A1, each operator corresponding 
to vertexes $\alpha_F$
and edges $(\alpha_F,L\ {\rm or}\ R)$ acts 
on $f$ and $g$ independently. 
Thus $w_{\alpha_F}\sim w_1^{n_{\alpha_F}}$. 
Secondly we consider the graph which consists of four vertexes
corresponding to $\alpha _A$, $\alpha _F$, and ``$L$''and ``$R$'' as
shown in Fig.~\ref{graph0.eps}. 
We also assume that one edge of the vertex corresponding to $\alpha _A$
connects with a vertex corresponding to $\alpha _F$. 
In this case, from additional diagram rule A3, another edge of the
vertex has to connect with ``$L$'' or ``$R$''. 
Since we can exchange the role ``$R$'' and ``$L$'' by the variable
transformation $p\mapsto 1-p$, ($p\in H_+$), 
we assume that one edge of the vertex corresponding to 
$\alpha _A$ connect with ``$L$''. 
The weight $w_{\Gamma }$ in this case is given by Eq.~(\ref{w2}), i.e.,
the integrals for the weight is given by replacing coordinate of the
vertex corresponding to $\alpha _F$ with coordinate of ``$R$'' in $H_+$. 
This result can be expanded to every graph 
though a graph includes the vertices $\alpha_0$. 
For example, we illustrate the calculation of a six vertices graph, 
which only includes $\alpha_0$ and $\alpha_F$, 
in Fig.~\ref{GD2}. 
At first we make the cluster having 
only vertices $\alpha_F$, $f$ and $g$ (fig.~\ref{GD2}(b)), 
which is corresponding to the following operator:
$ \frac{w^{n_{\alpha_F}}}{n_{\alpha_F}!}
\alpha_F^{i_1i_2}\alpha_F^{i_3i_4}(\partial_{p_{i_1}}\partial_{p_{i_3}}f)
(\partial_{p_{i_2}}\partial_{p_{i_4}}g)$. 
The edges from the vertices act on the cluster independently 
(fig.~\ref{GD2}(c)); we obtain the following operator:
$
\frac{w^{n_{\alpha_0}}}{n_{\alpha_0}!}\frac{w^{n_{\alpha_F}}}{n_{\alpha_F}!}
\alpha^{j_1j_2}_0\alpha^{j_3j_4}_0(\partial_{X^{j_1}}\alpha_F^{i_1i_2})
(\partial_{X^{j_3}}\alpha_F^{i_3i_4})(\partial_{X^{j_2}}\partial_{p_{i_1}}\partial_{p_{i_3}}f)
(\partial_{X^{j_4}}\partial_{p_{i_2}}\partial_{p_{i_4}}g)
$. 

The position of each vertex corresponding to $\alpha _A$ and
$\alpha _F$ can be move independently in integrals, and the entangled
integral does not appear. 
Therefore the weight $w_{n_{\alpha _A}}$ of a graph 
$\Gamma _{\alpha_A}\sim w_1^{n_{\alpha_A}}$ only depends on 
the number of vertexes
corresponding to $\alpha _A$ and $\alpha _F$, 
and $w_\Gamma =w_{n _{\alpha _A}} \cdot w_{n _{\alpha _F}}$
holds generally. 
From additional rule A2, we can similarly discuss about a graph 
$\Gamma_{\alpha_0}$, and obtain $w_{n_{\alpha_0}}\sim w_1^{n_{\alpha_0}}$. 
Finally, we can count the combination of $n_{\alpha_0}$,
$n_{\alpha_A}$ and $n_{\alpha_F}$, and it is given by
$\frac{(n_{\alpha_0}+n_{\alpha_A}+n_{\alpha_F})!}
{(n_{\alpha_A}+n_{\alpha_F})!n_{\alpha_0}!}
\cdot\frac{(n_{\alpha_A}+n_{\alpha_F})!}{n_{\alpha_A}!n_{\alpha_F}!}$.
Therefore we obtain the Eq: $w_{\Gamma }B_{\Gamma ,\alpha }=w_{n _{\alpha
_0}}B_{\Gamma _{\alpha _0},\alpha _0}\cdot
w_{n_{\alpha_A}}B_{\Gamma_{\alpha_A},\alpha_A}\cdot w_{n
_{\alpha _F}}B_{\Gamma _{\alpha _F},\alpha _F}$. 

The summation of each graph is easy, and we can derive the star product: 
$f\star g=\mu\circ{\cal F}_A(f\otimes g)$, where twisted element ${\cal F}_A$
is written as follow:
\begin{eqnarray}
{\cal F}_A&=&{\rm exp}\left\{\frac{{\rm i}\hbar}{2}
\left(\partial_{X^\mu}\otimes\partial_{p_\mu}
-\partial_{p\mu}\otimes\partial_{X^\mu}\right)\right\}\nonumber\\
&&\circ\ {\rm exp}\left\{\frac{{\rm i}\hbar}{2}\varepsilon^{ijk}s^k
\left(A_\mu^i\partial_{p_\mu}\otimes\partial_{s^j}
+\partial_{s^i}\otimes\partial_{s^j}-
\partial_{s^j}\otimes
A_\mu^i\otimes\partial_{p_\mu}\right)\right\}
\nonumber\\
&&\circ\ {\rm exp}
\left\{\frac{{\rm i}\hbar}{2}\left(F_{\mu\nu}^as^a+
F_{\mu\nu}\right)\partial_{p_\mu}\otimes\partial_{p_\nu}
\right\}. 
\end{eqnarray}

Because the action $I_A$ including a global U(1)$\times$SU(2) gauge
field, the action $I_A$ is
written as $I_A={\cal F}_A\circ{\cal F}_0^{-1}I_0$, 
thus, the map ${\cal F}_{0\mapsto A}:I_0\mapsto I_A$ is given by 
${\cal F}_{0\mapsto A}={\cal F}_A\circ{\cal F}_0^{-1}$, i.e., 
\begin{eqnarray}
 {\cal F}_{(0\mapsto A)}&=&{\rm exp}\left\{\frac{\mbox{i}\hbar}{2}
\left(\partial_{X^\mu}\otimes\partial_{p_\mu}
-\partial_{p\mu}\otimes\partial_{X^\mu}\right)\right\}\nonumber\\
&&\circ\ {\rm exp}\left\{\frac{\mbox{i}\hbar}{2}\varepsilon^{ijk}s^k
\left(A_\mu^i\partial_{p_\mu}\otimes\partial_{s^j}
+\partial_{s^i}\otimes\partial_{s^j}-
\partial_{s^j}\otimes
A_\mu^i\otimes\partial_{p_\mu}\right)\right\}
\nonumber\\
&&\circ\ {\rm exp}
\left\{\frac{\mbox{i}\hbar}{2}\left(F_{\mu\nu}^as^a+
F_{\mu\nu}\right)\partial_{p_\mu}\otimes\partial_{p_\nu}
\right\}\circ {\rm exp}\left\{
-\frac{\mbox{i}\hbar}{2}\varepsilon^{ijk}s^k\partial_{s^i}\otimes\partial_{s^j}
\right\}\nonumber\\
&&\circ\ {\rm exp}\left\{-\frac{\mbox{i}\hbar}{2}
\left(\partial_{X^\mu}\otimes\partial_{p_\mu}-
\partial_{X^\mu}\otimes\partial_{p_\mu}
\right)\right\}.\label{eq:F_02A}
\end{eqnarray}
The inverse map is given by the replacement of i by $-\mbox{i}$ in
the map (\ref{eq:F_02A}). 

\section{Twisted spin}\label{section:TS}
In this section, we will derive the twisted spin density, 
which corresponds to the spin density in commutative 
spacetime without the background SU(2) gauge field. 

First, we will derive a general form of the twisted spin current in
section \ref{subsec:DTSWS}, 
which is written by using the twisted variation operator. 
This operator is constituted of the coproduct and twisted element; 
the coproduct reflects the action rule 
of the global SU(2) gauge symmetry generator, 
and the twisted element 
represents gauge structure of the background gauge fields. 

In section \ref{sub:RDM}, at first, 
we will calculate the twisted spin density of the 
so-called Rashba-Dresselhaus model in the Wigner representation
using the general form of the twisted spin density, and next, 
we will find the twisted spin operator in real spacetime using 
correspondence between operators in commutative spacetime and
noncommutative phasespace.

\subsection{Derivation of a twisted spin in Wigner space}\label{subsec:DTSWS}
The Lagrangian density in the Wigner space is given by
\begin{equation}
 {\cal L}(X,p)=\left(p_0-\frac{\bm{p}^2}{2m}\right)
\star \left(\psi\psi^\dagger\right),
\end{equation}
where $m$ is the electric mass, 
$f\star g:=\mu\circ {\cal F}_A(f\otimes g)$ 
for any functions $f$ and $g$. 

The variation corresponding to the infinitesimal global SU(2) 
gauge transformation is defined as
\begin{eqnarray}
 \delta_{s^a}\psi&=&{\rm i}\vartheta s^a\psi,\\
 \delta_{s^a}\psi^\dagger&=&-{\rm i}\psi^\dagger s^a,\\
 \delta_{s^a}x^\mu&=&0,
\end{eqnarray}
where $\theta$ represents an infinitesimal parameter. 
Therefore, the variation of the Lagrangian density 
${\cal L}(X,p):=\hat{\cal L}\star \psi\psi^\dagger$
is given by
\begin{equation}
 \delta_{s^a}\left({\cal L}(X,p)\right):=
\hat{\cal L}\star{\rm i}\theta s^a\star \psi\psi^\dagger
-\hat{\cal L}\star\psi\psi^\dagger\star{\rm i}\theta s^a
\label{eq:dL}
\end{equation}
Here we introduce the Grassmann numbers $\lambda_{1,2,3}$ 
($\lambda_3:=\lambda_1\lambda_2$), 
the product $\mu$, and the coproduct $\triangle_\eta$, where 
the coproduct satisfies
\begin{eqnarray}
 \triangle_\eta(f):=f\otimes \eta+\eta\otimes f
\end{eqnarray}
for any functions $f$ and operator $\eta$. 
The equation (\ref{eq:dL}) can be rewritten as
\begin{eqnarray}
 \delta_{s^a}\left({\cal L}\right)&=&
\int d\lambda_3\mu\circ{\cal F}_A\left(
\hat{\cal L}\otimes\left(
{\rm i}\lambda_1\theta s^a\star\lambda_2\psi\psi^\dagger
+\lambda_2\psi\psi^\dagger\star
{\rm i}\lambda_1\theta s^a\right)\right)\nonumber\\
&=&\int d\lambda_3\mu\circ{\cal F}_A\left(
\hat{\cal L}\otimes\mu\circ{\cal F}_A\triangle_{{\rm i}\lambda_1\theta s^a}
\right)\nonumber\\
&=&\int d\lambda_3\mu\circ {\cal F}_A({\rm id\otimes \mu})
\circ ({\rm id}\otimes{\cal F}_A)\circ({\rm id}\otimes 
\triangle_{{\rm i}\lambda_1\theta s^a})\circ
(\hat{\cal L}\otimes \lambda_2\psi\psi^\dagger)\nonumber\\
&=&{\rm i}\int d\lambda_3\mu\circ({\rm id}\otimes \mu)\circ
({\rm id}\otimes\triangle){\cal F}_A
\circ({\rm id}\otimes {\cal F}_A)\nonumber\\
&&\circ({\rm id}\circ\theta^{1/2}\otimes\theta^{1/2})
\circ({\rm id}\otimes\triangle_{\lambda_1s^a})
\circ(\hat{\cal L}\otimes\lambda_2\psi\psi^\dagger),
\label{eq:dL2}
\end{eqnarray}
where we used $\int d\lambda_i\lambda_j=\delta_{ij}
(i,j=1,2,3)$ with the Kronecker delta $\delta_{ij}$
Here we introduce the following symbols:
\begin{eqnarray}
 \hat{\mu}&:=&\mu\circ({\rm id}\otimes\mu),\\
 \hat{\cal F}_A&:=&({\rm id}\otimes\triangle)
{\cal F}_A\circ({\rm id}\otimes{\cal F}_A),\\
\hat{\theta}&:=&\theta^{1/2}\otimes\theta^{1/2},\\
\hat{\triangle}_{\lambda_1s^a}&:=&
({\rm id}\otimes\triangle_{\lambda_1s^a}),
\end{eqnarray}
where the coproduct in the differential operator space 
is defined as
\begin{equation}
 \triangle(d_n):=
\sum_{\substack{i+j=n \\ i\geq 0,\ j\geq 0}}
d_i\otimes d_j.
\end{equation}
Here vectors $\{d_0,d_1,\dots\}$ 
corresponding to following operators:
$d_0:={\rm id}$ and $d_n:=(1/n!)(\partial^n/\partial p^n)$, 
or $d_0:={\rm id}$ and 
$d_n:=(1/n!)(\partial^n/\partial x^n)$
($n=1,2,\dots$), 
and $d_l$ are bases of a vector space 
$B(k):=\bigoplus_{l=0}^\infty kd_l$
($k$ represents a scalar). 
The coproduct $\triangle$ in the vector space $B(k)$ satisfies 
the coassociation law: 
$(\triangle\otimes{\rm id})\otimes\triangle
=({\rm id}\otimes\triangle)\circ\triangle$  
because 
\begin{eqnarray}
 (\triangle\otimes{\rm id})\circ\triangle(d_n)&=&
\sum_{\substack{i+j=n \\ i\geq 0,\ j \geq 0}}
\triangle(d_i)\otimes d_j\nonumber\\
&=&
\sum_{\substack{i+j=n \\ i\geq 0,\ j\geq 0}}
\left(
\sum_{\substack{k+l=i \\ k\geq 0,\ l\geq 0}}
d_k\otimes d_l\otimes d_j
\right)\nonumber\\
&=&
\sum_{\substack{k+l+j=n \\ k\geq 0,\ l\geq 0,\ j\geq 0}}
d_k\otimes d_l\otimes d_j\label{eq:CoAssociativeLaw.L.DTS}
\end{eqnarray}
and
\begin{eqnarray}
 ({\rm id}\otimes\triangle)\circ\triangle(d_n)&=&
\sum_{\substack{i+j=n \\ i\geq 0,\ j\geq 0}}
d_i\otimes\triangle(d_j)\nonumber\\
&=&\sum_{\substack{i+k+l=n \\ i\geq 0,\ k\geq 0,\ l\geq 0}}
d_i\otimes d_k\otimes d_l.\label{eq:CoAssociativeLaw.R.DTS}
\end{eqnarray}
It represents the Leibniz rule with respect to 
the differential operator $\partial_\mu$. 
For instance, 
$({\rm id}\otimes\triangle)
(\partial_\mu\otimes\partial_\nu)
:=\partial_\mu\otimes\triangle(\partial_\nu)
=\partial_\mu\otimes\partial_\nu\otimes{\rm id}
+\partial_\mu\otimes{\rm id}\otimes\partial_\nu$; 
it corresponds to the following calculation: 
\begin{equation}
 \partial_\mu f\cdot\partial_\nu(g\cdot h)=\partial_\mu
  f\cdot\partial_\nu g\cdot h+\partial_\mu f\cdot g\cdot\partial_\nu h.
\end{equation}

The variation (\ref{eq:dL2}) is rewritten as
\begin{equation}
 \delta_{s^a}{\cal L}(X,p)=
{\rm i}\int d\lambda_3\hat{\mu}\circ
\hat{\cal F}_A\circ\hat{\theta}\circ
\hat{\triangle}_{\lambda_1s^a}\circ
(\hat{\cal L}\otimes \lambda_2\psi\psi^\dagger).
\label{eq:dL3}
\end{equation}

If we replace 
$\hat{\triangle}^t_{\lambda_1s^a}:=
\hat{\cal F}_A^{-1}\hat{\triangle}_{\lambda_1s^a}
{\cal F}_0$ 
with 
$\hat{\triangle}_{\lambda_1s^a}$ 
in equation (\ref{eq:dL3}), 
the integrals in terms of $x$ and $p$ 
of the right-hand side of 
Eq.~(\ref{eq:dL3}) become zero 
because 
$\hat{\cal F}_A\circ\hat{\triangle}^t_{\lambda_1s^a}$ 
does not include the SU(2) field, 
which breaks the global SU(2) gauge symmetry, 
in the case that the parameter $\theta$ is constant.  
Therefore, for the action 
$S:=\int d^{\rm Dim}Xd^{\rm Dim}p{\cal L}(X,p)/(2\pi\hbar)^{\rm Dim}$, 
\begin{equation}
 \delta_{s^a}^tS:=
{\rm tr}\iiint d\lambda_3d^{\rm Dim}X
\frac{d^{\rm Dim}p}{(2\pi\hbar)^{\rm Dim}}
\hat{\mu}\circ\hat{\cal F}_A\circ\hat{\theta}\circ
\hat{\triangle}^t_{\lambda_1s^a}(\hat{\cal L}\otimes
\lambda_2\psi\psi^\dagger)
\end{equation}
is the infinitesimal SU(2) gauge transformation with background 
SU(2) gauge fields.

Because $\delta^t_{s^a}S=0$, we can write
\begin{equation}
 \delta^t_{s^a}S=\int d^{\rm Dim}X\theta\left(
\partial^\mu j^t_\mu\right).
\end{equation}
In the case that the infinitesimal parameter 
depends on the spacetime coordinate, 
this equation can be written as
\begin{eqnarray}
 \delta^t_{s^a}S&=&\int d^{\rm Dim}X\theta(X)\left(
\partial^\mu j^t_\mu\right)\nonumber\\
&=&-\int d^{\rm Dim}X\left(
\frac{\partial\theta(X^\prime)}{\partial X^\prime_\mu}
\right)
j^t_\mu(X^\prime).
\end{eqnarray}
Therefore, we obtain the twisted Noether current
\begin{equation}
 j^t_\mu=-\frac{\delta^t_{s^a}S}{\delta(\partial_\mu\theta(X))}.
\end{equation}

In particular, the twisted spin 
\begin{eqnarray}
 S^t_a&=&\int d\bm{X}j^t_0\nonumber\\
 &=&\int d\bm{X}\frac{\delta^t_{s^a}S}{\delta(\partial_T\theta)}
\end{eqnarray}
is conserved quantity. 
Here, we assumed that the SU(2) gauge is static one. 
However, we do not use this condition in the derivation of 
the twisted Noether charge and current density. 
Then, we can derive the virtual twisted spin with 
a time-dependent SU(2) gauge: $\tilde{S}^t_a$. 
In this case, we only use 
the time-dependent SU(2) gauge field strength 
$F^a_{\mu\nu}$, which has non-zero space-time components 
$F^a_{0i}$ ($i=1,2,\dots,{\rm Dim}-1$).

Here, we discuss the adiabaticity of the twisted spin. 
In the case that SU(2) gauge fields have time dependence, 
the twisted spin is not conserved. 
Now, we assume that $A^a_\mu=\lambda(t)C^a_\mu$ ($a=x,y,z$) 
with constant fields $C^a_\mu=(0,\bm{C}^a)$; 
$\lambda(t)$ is an adequate slowly function dependent on time. 
Because $\tilde{S}^t_a$ includes 
$F_{\mu\nu}^{-1}\sim\frac{1}{(1+(\dot{\lambda})^2\bm{C}\cdot\bm{C})}
\left(
\begin{array}{cc}
0 & \dot{\lambda}\bm{C} \\
-\dot{\lambda}\bm{C} & \lambda^{-2}[\bm{C}_i,\bm{C}_j]^{-1}
\end{array}
\right)$ 
with $\dot{\lambda}\equiv d\lambda/dt$, 
the difference between $\tilde{S}^t_a$ and $S^t_a$ comes from only that 
between inverse of field strength: 
$\Delta F^{-1}\sim
\frac{1}{1+(\dot{\lambda})^2}\lambda^{-2}[C_i,C_j]^{-1}
-\lambda^{-2}[C_i,C_j]^{-1}\sim 
(\dot{\lambda}/\lambda)^2[C_i,C_j]^{-1}$. 
Therefore we obtain 
\begin{equation}
 \frac{dS^t_a}{dt}={\cal O}(\dot{\lambda}^2).
\end{equation}
This means that $S^t_a$ is the adiabatic invariance. 
Namely, for the infinitely slow change in $\lambda(t)$ during the 
time period $T(\rightarrow\infty)$, 
$S^t_a$ remains constant while $\Delta\lambda=\lambda(T)-\lambda(0)$
is finite. 
This fact is essential for the spin-orbit echo proposed in
\cite{Sugimoto2}.

\subsection{Rashba-Dresselhaus model}\label{sub:RDM}
Here we apply the formalism developed so far to an explicit model, i.e.,
the so-called Rashba-Dresselhaus model given by
\begin{equation}
 H=\frac{\hat{\bm{p}}^2}{2m}
+\alpha(\hat{p}_x\hat{\sigma}_y-\hat{p}_y\hat{\sigma}_x)
+\beta(\hat{p}_x\hat{\sigma}_x-{\hat p}_y\hat{\sigma}_y)
+V(\hat{x})
\end{equation}
with a potential $V(\hat{x})$, 
where $\alpha$ and $\beta$ are the Rashba and Dresselhaus parameters, 
respectively. 
Completing square in terms of $\hat{p}$, 
we obtain 
$A^x_x=-2m\beta/(\hbar q)$, 
$A^y_x=-2m\alpha/(\hbar q)$, 
$A^x_y=2m\alpha/(\hbar q)$, 
$A^y_y=2m\beta/(\hbar q)$, 
$A_0=m(\alpha^2+\beta^2)/e$,
and $A^z_{x,y}=A^{x,y}_z=A^z_z=A_0^{x,y,z}=A_{x,y,z}=0$, 
where $q=|e|/(mc^2)$. 

To calculate the twisted symmetry generator
$\hat{\triangle}^t(\lambda_2s^a)$, 
we first consider the $\hat{{\cal F}}_{0}$. 
$({\rm id}\otimes\triangle){\cal F}_{0}$ is given by
\begin{eqnarray}
 ({\rm id}\otimes\triangle){\cal F}_{0}&=&
\exp\left\{
\frac{{\rm i}\hbar}{2}\Big(
 \partial_{X^\mu}\otimes 1\otimes\partial_{p_\mu}
+\partial_{X^\mu}\otimes\partial_{p_\mu}\otimes 1
-\partial_{p_\mu}\otimes 1\otimes\partial_{X^\mu}
-\partial_{p_\mu}\otimes\partial_{X^\mu}\otimes 1
\Big)\right.\nonumber\\
&&\left.
+\frac{i}{2}\varepsilon^{ijk}\Big(
 s_k\partial_{s_i}\otimes 1\otimes\partial_{s_j}
+s_k\partial_{s_i}\otimes\partial_{s_j}\otimes 1
\Big)\right\},
\end{eqnarray}
and $\hat{\cal F}_{0}$ is given by
\begin{eqnarray}
 \hat{\cal F}_{0}&=&\exp\left\{
\frac{{\rm i}\hbar}{2}\Big(
 \partial_{X^\mu}\otimes 1\otimes\partial_{p_\mu}
+\partial_{X^\mu}\otimes\partial_{p_\mu}\otimes 1
-\partial_{p_\mu}\otimes 1\otimes\partial_{X^\mu}
-\partial_{p_\mu}\otimes\partial_{X^\mu}\otimes 1
\Big)\right.\nonumber\\
 &&\left.
+\frac{i}{2}\varepsilon^{ijk}\Big(
 s_k\partial_{s_i}\otimes 1\otimes\partial_{s_j}
+s_k\partial_{s_i}\otimes\partial_{s_j}\otimes 1
\Big)\right\}\nonumber\\
 &&\circ\exp\left\{
\frac{{\rm i}\hbar}{2}\Big(
 1\otimes\partial_{X^\mu}\otimes\partial_{p_\mu}
-1\otimes\partial_{p_\mu}\otimes\partial_{X^\mu}
\Big)\right.\nonumber\\
 &&\left.
+\frac{i}{2}\varepsilon^{ijk}\Big(
 1\otimes s_k\partial_{s_i}\otimes\partial_{s_j}
\Big)\right\}\nonumber\\
 &=&\exp\left\{
\frac{{\rm i}\hbar}{2}\Big(
 1\otimes \partial_{X^\mu}\otimes\partial_{p_\mu}
+\partial_{X^\mu}\otimes 1\otimes\partial_{p_\mu}
+\partial_{X^\mu}\otimes\partial_{p_\mu}\otimes 1
\right.\nonumber\\
&&\left.
-1\otimes \partial_{p_\mu}\otimes\partial_{X^\mu}
-\partial_{p_\mu}\otimes 1\otimes\partial_{X^\mu}
-\partial_{p_\mu}\otimes\partial_{X^\mu}\otimes 1
\Big)\right.\nonumber\\
 &&\left.
+\frac{i}{2}\varepsilon^{ijk}\Big(
 1\otimes s_k\partial_{s_i}\otimes\partial_{s_j}
+s_k\partial_{s_i}\otimes 1\otimes\partial_{s_j}
+s_k\partial_{s_i}\otimes\partial_{s_j}\otimes 1
\Big)\right\}\nonumber\\
&\equiv&{\cal G}_0.
\end{eqnarray}
Similarly, $\hat{\cal F}_A$ is given by
\begin{eqnarray}
 \hat{\cal F}_A
 &=&\exp\left\{
\frac{{\rm i}\hbar}{2}\Big(
 1\otimes \partial_{X^\mu}\otimes\partial_{p_\mu}
+\partial_{X^\mu}\otimes 1\otimes\partial_{p_\mu}
+\partial_{X^\mu}\otimes\partial_{p_\mu}\otimes 1
\right.\nonumber\\
&&\left.
-1\otimes \partial_{p_\mu}\otimes\partial_{X^\mu}
-\partial_{p_\mu}\otimes 1\otimes\partial_{X^\mu}
-\partial_{p_\mu}\otimes\partial_{X^\mu}\otimes 1
\Big)\right\}\nonumber\\
 &&\circ
\exp\left\{\frac{i}{2}\varepsilon^{ijk}\Big(
 1\otimes s_k\partial_{s_i}\otimes\partial_{s_j}
+s_k\partial_{s_i}\otimes 1\otimes\partial_{s_j}
+s_k\partial_{s_i}\otimes\partial_{s_j}\otimes 1
\right.\nonumber\\
&&
+1\otimes s^kA^i_\mu\partial_{p_\mu}\otimes\partial_{s^j}
+s^kA^i_\mu\partial_{p_\mu}\otimes 1\otimes\partial_{s^j}
+s^kA^i_\mu\partial_{p_\mu}\otimes\partial_{s^j}\otimes 1
\nonumber\\
&&\left.
-1 \otimes\partial_{s^j}\otimes s^kA^i_\mu\partial_{p_\mu}
-\partial_{s^j}\otimes 1\otimes s^kA^i_\mu\partial_{p_\mu}
-\partial_{s^j}\otimes s^kA^i_\mu\partial_{p_\mu}\otimes 1
\Big)\right\}\nonumber\\
&&\circ\exp\left\{
\frac{{\rm i}\hbar}{2}\Big(
 1\otimes \hat{F}_{\mu\nu}\partial_{p_\mu}\otimes\partial_{p_\nu}
+\hat{F}_{\mu\nu}\partial_{p_\mu}\otimes 1\otimes\partial_{p_\nu}
+\hat{F}_{\mu\nu}\partial_{p_\mu}\otimes\partial_{p_\nu}\otimes 1
\Big)\right\}\nonumber\\
&\equiv&{\cal G}^A_{Xp}\circ{\cal G}^A_{sp}\circ{\cal G}^A_{pp}.
\end{eqnarray}
We note that the operators 
${\cal G}_0~{\rm and}~{\cal G}^A_{Xp,sp,pp}$ have
each inverse operator, which are denoted by
$\overline{\cal G}_0~{\rm and}~\overline{\cal G}^A_{Xp,sp,pp}$, 
respectively. 
Here, the overline $-$ represents the complex conjugate. 

Because the twisted variation is $\hat{\mu}\circ\hat{\cal F}_A\circ
\hat{\theta}\circ \hat{\cal F}^{-1}_A\circ\hat{\triangle}_{\lambda_1s^a}
\circ {\cal F}_0({\cal L}\otimes \lambda_2\psi\psi^\dagger)
$, the infinitesimal parameter $\hat{\theta}$ becomes an operator 
$\hat{\cal F}_A\circ
\hat{\theta}\circ \hat{\cal F}^{-1}_A$. 
It is calculated by using the operator formula
\begin{equation}
 e^BCe^{-B}=\sum_{n=0}^\infty
\frac{1}{n!}\underbrace{[B,[B,\cdots[B,}_nC]\cdots]]
\end{equation}
for any operators $B$ and $C$. 
In the calculation, 
one will use the following formula in midstream:
\begin{eqnarray}
 \sum_{l=0}\frac{D^l}{(l+1)!}&=&\int_0^1d\lambda e^{\lambda D},\\
 \sum_{l=0}\frac{D^l}{(l+2)!}&=&\int_0^1d\lambda\int_0^\lambda
d\lambda^\prime e^{\lambda^\prime D},
\end{eqnarray}
and so on, for any operator $D$.

From these results of the calculations, 
we obtain the twisted spin as follow
\begin{equation}
 s^t_a=\mu\circ{\cal F}_A\circ(s^a\otimes{\rm id})
\circ\tilde{\Upsilon}\circ({\rm id}\otimes G)
\circ\tilde{\Upsilon}^\dagger,
\end{equation}
where $\tilde{\Upsilon}$ is defined as
\begin{eqnarray}
 \tilde{\Upsilon}&:=&
\frac{1}{2}\left[
e^{\frac{{\rm i}}{2}(\alpha+\beta)\sigma_-
\big(\partial_{p_x}+\partial_{p_y}\big)}
\otimes 
e^{\frac{{\rm i}}{2}(\alpha-\beta)\sigma_+
\big(\partial_{p_x}-\partial_{p_y}\big)}
\right]\circ\left[
e^{-\frac{{\rm i}}{2}\big(
{\partial_X}\otimes\partial_Y+
{\partial_Y}\otimes\partial_X\Big)
\circ\big(\frac{1}{2m(\alpha^2-\beta^2)}
\otimes \sigma^z\big)
}\right.\nonumber\\
&&\left.
+e^{-\frac{{\rm i}}{2}\big(
{\partial_X}\otimes\partial_Y+
{\partial_Y}\otimes\partial_X\Big)
\circ\big(
\sigma^z\otimes\frac{1}{2m(\alpha^2-\beta^2)} 
\big)
}\right]
+e^{\frac{{\rm i}}{2}(\alpha+\beta)\sigma_-
\big(\partial_{p_x}+\partial_{p_y}\big)}
\otimes \sin^2\left(
\frac{{\rm i}\hbar
\big(\partial_X-\partial_Y\big)}{8\sqrt{2}m(\alpha+\beta)}
\right),
\end{eqnarray}
where $\sigma_\pm:=\sigma_x\pm\sigma_y$.

Finally, we will rewrite the twisted spin 
as an operator form in commutative spacetime. 
Roughly speaking, the operator in the commutative spacetime.
and the one in the noncommutative Wigner space 
have the following relations 
(the left-hand side represents operators on the Wigner space; 
the right-hand side represents operators on the commutative spacetime):
\begin{eqnarray}
 X^\mu\star &\Leftrightarrow& \hat{x}^\mu,\\
 p_\mu\star &\Leftrightarrow& \hat{p}_\mu,\\
 s^a\star&\Leftrightarrow&\hat{s}^a,\\
 {\rm i}\hbar\partial_{p_\mu}\star&\Leftrightarrow&\hat{x}^\mu,\\
 -{\rm i}\hbar\partial_{X^\mu}\star&\Leftrightarrow&\hat{p}_\mu,
\end{eqnarray}
because $[X^\mu,p_\nu]_\star:=X^\mu\star p_\nu-p_\nu\star X^\mu=
{\rm i}\hbar\delta_\nu^\mu$ is equal to the commutation of 
the operator form: 
$[\hat{x}^\mu,\hat{p}_\nu]={\rm i}\hbar\delta_\nu^\mu$. 
The equivalence of $s^t_a$ and 
the twisted spin in the operator form 
on the commutative spacetime $\hat{s}^t_a$ 
can be confirmed using the Wigner transformation
in terms of $\hat{s}^t_a$. 

The operator form of the twisted spin is given by
\begin{eqnarray}
 s^t_a=\frac{\hbar}{2}\psi^\dagger\Upsilon^\dagger\sigma^a\Upsilon\psi,
\label{eq:tso}
\end{eqnarray}
where 
\begin{eqnarray}
 \Upsilon&=&\lim_{x^\prime\rightarrow x}
\frac{1}{2}e^{\frac{{\rm i}}{2}m(\alpha+\beta)\sigma_-x_+}
\left[
e^{-\frac{{\rm i}}{2}m(\alpha-\beta)\sigma_+x^\prime_-}
e^{-\frac{{\rm i}}{2}
\frac{\sigma^z}{2m^2(\alpha^2-\beta^2)}
\big(\overleftarrow{\partial}_{x^\prime}
\partial_y+
\overleftarrow{\partial}_{y^\prime}
\partial_x\big)}\right.\nonumber\\
&&\left.+e^{-\frac{{\rm i}}{2}
\frac{\sigma^z}{2m^2(\alpha^2-\beta^2)}
\big(\partial_{x^\prime}
\partial_y+
\partial_{y^\prime}
\partial_x\big)}
e^{-\frac{{\rm i}}{2}m(\alpha-\beta)\sigma_+x^\prime_-}
+2\sin^2\left(
\frac{{\rm i}\hbar(\partial_x-\partial_y)}
{8\sqrt{2}m(\alpha+\beta)}
\right)\right]
\end{eqnarray}
with $x_\pm:=x\pm y$. 

This operator in Eq.~(\ref{eq:tso}), when integrated over $\bm{X}$, 
is the conserved quantity for any potential configuration 
$V(\hat{x})$ as long as $\alpha$, $\beta$ 
are static and the electron-electron interaction is neglected. 

\section{Conclusions}\label{sec:Conc}

In this paper, we have derived the conservation of the
twisted spin and spin current densities. 
Also the adiabatic invariant nature of the 
total twisted spin integrated over the space is shown. 
Here we remark about the limit of validity of this conservation law.
First, we neglected the dynamics of the electromagnetic field 
$A_\mu$ which leads to the electron-electron interaction. 
This leads to the inelastic electron scattering, which is not 
included in the present analysis, and most likely 
gives rise to the spin relaxation. This inelastic scattering
causes the energy relaxation and hence the memory of 
the spin will be totally lost after the inelastic lifetime.
This situation is analogous to the two relaxation times
$T_1$ and $T_2$ in spin echo in NMR and ESR.
Namely, the phase relaxation time $T_2$ is
usually much shorter than the energy relaxation 
time $T_1$, and the spin echo is possible 
for $T< T_1$. Similar story applies to spin-orbit echo~\cite{Sugimoto2}
where the recovery of the spin moment is possible only within 
the inelastic lifetime of the spins. 
However, the generalization of the present study to the
dynamical $A_\mu$ is a difficult but important issue left 
for future investigations. Also the effect of the higher order
terms in $1/(mc^2)$ in the derivation of the effective 
Lagrangian from Dirac theory requires to be scrutinized. 

Another direction is to explore the twisted conserved
quantities in the non-equilibrium states. Under the static electric field,
the system is usually in the current flowing steady state. 
Usually this situation 
is described by the linear response theory, 
but the far from equilibrium states
can in principle be described by the non-commutative 
geometry~\cite{SOnoda,Sugimoto}. 
The nonperturbative effects
in this non-equilibrium states are the challenge for theories, and deserve
the further investigations.  

The author thanks Y.S. Wu, F.C. Zhang, K. Richter, V. Krueckl, 
J. Nitta, and S. Onoda for useful discussions.
This work was supported by Priority Area Grants, 
Grant-in-Aids under the Grant number 21244054, 
Strategic International Cooperative Program (Joint Research Type) 
from Japan Science and Technology Agency, 
and by Funding Program for World-Leading Innovative 
R and D on Science and Technology (FIRST Program).

\end{document}